\documentclass[aps,pre,amsmath,amssymb,reprint,floatfix]{revtex4-1}
\usepackage{lipsum}
\usepackage{amssymb,euscript,latexsym,amsmath,amsthm}
\usepackage{graphicx}
\usepackage{color}
\usepackage{setspace}
\usepackage{float}
\usepackage{bm}
\usepackage[dvipsnames]{xcolor}
\usepackage{chngpage}
\usepackage{hyperref}
\usepackage{subcaption}
\usepackage{cancel}
\usepackage{setspace}
\usepackage{graphicx}

\usepackage[T1]{fontenc}
\usepackage[utf8]{inputenc}
\usepackage[font=small,labelfont=bf,tableposition=top]{caption}
\usepackage{booktabs,multirow}

\DeclareGraphicsExtensions{.eps,.ps,.pdf}

\def\pp#1#2{{\frac{\partial #1}{\partial #2}}}

\def\pp#1#2{{\frac{\partial #1}{\partial #2}}}

\def\dif{\mathrm{d}}

\def\f#1#2{{\frac{#1}{#2}}}
\def\*{\cdot}

\def\vect#1{\boldsymbol{\mathsf{#1}}}
\def\vect#1{\boldsymbol{\mathbf{#1}}}

\begin{document}

\title{Instability-driven Oscillations of Elastic Microfilaments}

\author{Feng Ling}
\author{Hanliang Guo}
\author{Eva Kanso}
\email{kanso@usc.edu}
\affiliation{Department of Aerospace and Mechanical Engineering,  \\ University of Southern California, Los Angeles, California 90089, USA}

\date{\today}
\begin{abstract}
Cilia and flagella are highly conserved slender organelles that exhibit a variety of rhythmic beating patterns from non-planar cone-like motions to planar wave-like deformations. Although their internal structure, composed of a microtubule-based axoneme driven by dynein motors, is known, the mechanism responsible for these beating patterns remains elusive. Existing theories suggest that the dynein activity is dynamically regulated, via a geometric feedback from the cilium's mechanical deformation to the dynein force. An alternative, open-loop mechanism based on a `flutter' instability was recently proven to lead to planar oscillations of elastic filaments under follower forces.
Here, we show that an elastic filament in viscous fluid, clamped at one end and acted on by an external distribution of compressive axial forces, exhibits a Hopf bifurcation
that leads to non-planar spinning of the buckled filament at a locked curvature.
We also show the existence of a second bifurcation, at larger force values, that induces a transition from  non-planar spinning to planar wave-like oscillations.
We elucidate the nature of these instabilities using a combination of nonlinear numerical analysis, linear stability theory, and low-order bead-spring models. 
Our results show that away from the transition thresholds, these beating patterns are robust to perturbations in the distribution of axial forces and in the filament configuration.
These findings support the theory that an open-loop, instability-driven mechanism could explain both the sustained oscillations and the wide variety of periodic beating patterns observed in cilia and flagella. 
\end{abstract}
\maketitle

\section{Introduction}
\label{sec:intro}

\begin{figure}[!t]
	\centering
	\includegraphics[scale=2.67]{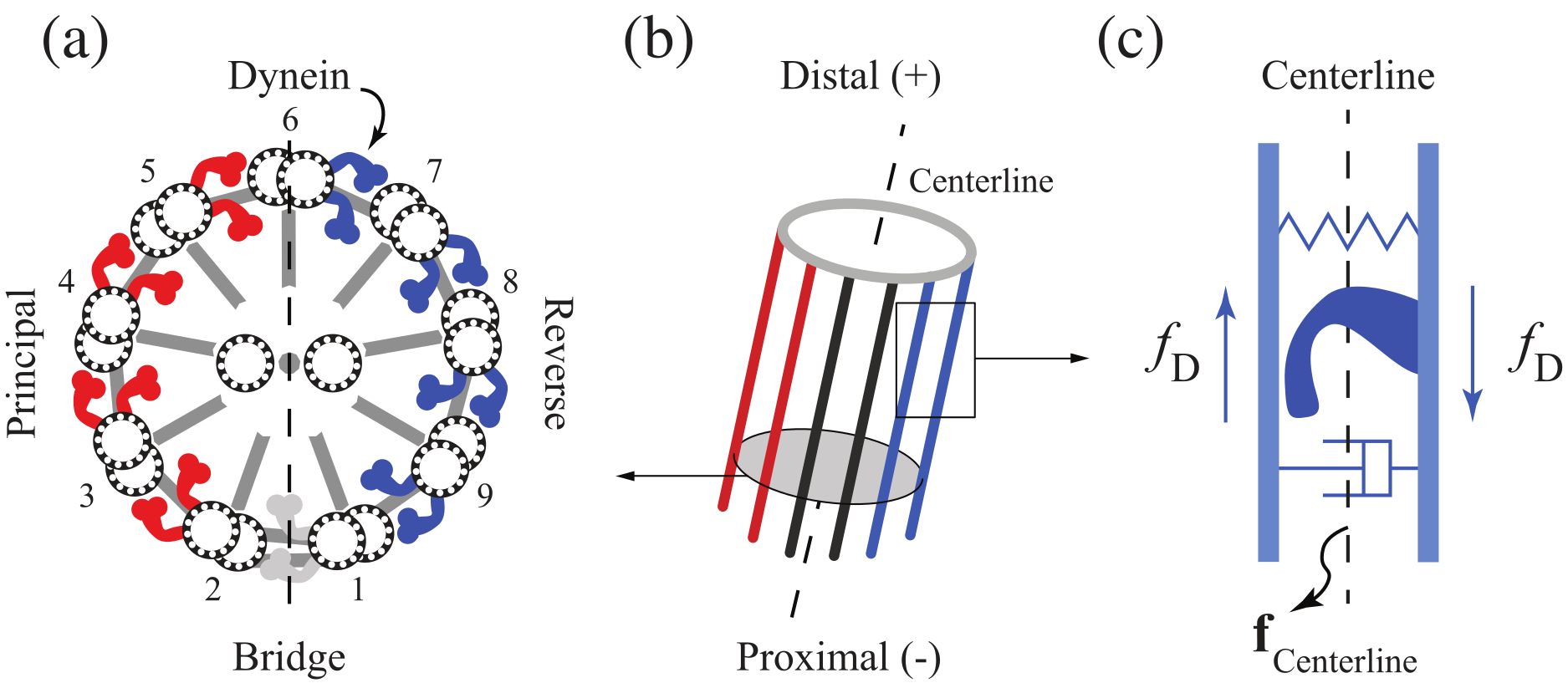}
	\caption{
	\textbf{Internal structure of a cilium/flagellum:} (a) represents the cross section of a typical cilium/flagellum. A bridge exists between two of the doublets, thus the Dynein motors are usually separated into principal (P) and reverse (R) sides. Feedback control hypotheses \cite{Brokaw1971,Lindemann1994b,Sartori2016a} are based on the theory that P,R sides are activated in alternating fashion due to kinematic constraints. But even when both sides are activated in roughly equal amounts, a flutter instability could drive the cilium into oscillations \cite{Bayly2016,Hu2018}. (b-c) In both theories, internal force couples ($f_D$) generated by dynein motors walking towards the minus (-) end of microtubule doublets effectively apply an ``external force'' on the centerline. This external loading is \textit{explicitly} coupled with the kinematics of the centerline.}
	\label{fig:cilia}
\end{figure}

Cilia and flagella are microscopic hair-like organelles found in many eukaryotic cells, from
single-celled protozoa to mammalian epithelial surfaces. They are driven into oscillatory motion by an intricate internal structure, referred to as the central axoneme, composed of microtubule doublets and dynein molecular motors (see figure~\ref{fig:cilia}).  
Despite the highly-conserved structure of the central axoneme across eukaryotic cells, cilia and flagella of different cells exhibit wildly different beating patterns, from non-planar, cone-like, motions to planar, wave-like, deformations \cite{Sleigh1968,Brennen1977}. 
The mechanisms that regulate the activity of the dynein motors, causing them to produce oscillatory motions, remain elusive, and even less is known about 
the mechanisms leading to this diversity in beating patterns~\cite{Blake1972,Chwang1971,Hirokawa2006,Meng2014}.
Several theoretical studies support the hypothesis that dynein motors are regulated by a geometric feedback from  mechanical deformations to molecular activity~\cite{Brokaw1971, Brokaw2009, Riedel-Kruse2007, Sartori2016a}. 

Brokaw was the first to show that delayed feedback from curvature to dynein activity could lead to oscillations~\cite{Brokaw1971,Brokaw1972,Brokaw1975,Brokaw1985}. Hines and Blum developed detailed models of elastic filaments that generate sustained oscillations with curvature feedback control~\cite{Hines1983}.
Murase, J\"{u}licher and others demonstrated the existence of oscillatory modes in sliding-control models~\cite{Murase1989,Hilfinger2009a,Hilfinger2009b,Camalet2000}. Lindemann proposed a `geometric clutch' hypothesis, where the dynein activity changes as a function of the spacing between doublet pairs leading to axoneme oscillations~\cite{Lindemann1994a,Lindemann1994b,Lindemann2002,Bayly2014}. 
A comparison of these various feedback mechanisms to experimental observations seems to favor the hypothesis of regulation by curvature feedback control~\cite{Sartori2016a}.

Although these feedback models are appealing, Bayly and Dutcher argued convincingly that evidence supporting the hypothesis that dynein regulation is required for bending oscillations remains circumstantial~\cite{Bayly2016}.
They then proposed an alternative mechanism that does not require regulation of dynein activity to generate oscillatory motions~\cite{Bayly2016,Hu2018}. The mechanism relies on a dynamic buckling instability induced by the \textit{internal} dynein forces, which apply axial stresses to the axoneme as it deforms. 
This instability is reminiscent to the classic Euler buckling instability, but instead of the familiar static instability under fixed forces, these follower forces lead to a dynamic buckling instability, known as a `flutter' instability in aeroelasticity~\cite{Dowell1989} or, more informally, as the `garden hose' instability~\cite{Paidoussis1998,Paidoussis2008}. 
To demonstrate that this dynamic instability could lead to sustained oscillations in cilia and flagella, Bayly and Dutcher used elaborate models of the axoneme, including finite element analysis and a model of two filaments, representing pairs of microtubule doublets, connected via passive elements of elastic spring and viscous damping. 
Han and Peskin also proposed an elaborate model of the axoneme structure that leads to sustained oscillations via a dynamic instability~\cite{Han2018}. 
De Canio, Lauga, and Goldstein explored the concept of instability-driven oscillations in the context of a simpler model, consisting of a single microfilament, confined to planar motion, clamped at one end and acted on by an \textit{external} follower force at the free end~\cite{DeCanio2017}.

Mathematical models, whether in support of curvature-feedback control or instability-driven oscillations, should be regarded as an abstraction of the intricate internal structure of cilia and flagella; see figure~\ref{fig:cilia}. 
Representations of the details of this structure require several assumptions in terms of model parameters (\textit{e.g.}, stiffness and damping coefficients for inter-doublet links),  internal geometric constraints, and dynein activation laws \cite{Bayly2014,Bayly2016,Han2018,Hu2018}.
Such level of detail is often ignored in favor of analytical rod models of the cilium centerline, where the internal dynein forces can be nominally treated as a distribution of ``external'' forces or moments along the centerline.
For example, in \cite{Eloy2012} and references therein, the internal forces are modeled in terms of a distribution of external moments only along the filament centerline (note that one could rewrite these moments in terms of tangential and normal forces). In~\cite{Camalet2000,Sartori2016a}, the cilium is reduced first to two coupled filaments, then to an effective set of equations governing the dynamics of the centerline of these two filaments subject to a distribution of active tangential and normal forces; see \cite[Eq.~(12)]{Sartori2016a}. 
Although it is clear from a homogenization approach that such centerline representation should exist,
deriving the equations of motion and effective forces acting on the centerline from more elaborate, coupled, multi-filament models of the axoneme structure is generally a difficult task; 
it typically leads to forces that are \textit{explicitly} coupled to the centerline curvature (as in~\cite[Eq.~(12)]{Sartori2016a} and~\cite[Eq.~(A18-19)]{Sartori2016b} for example). 
This coupling could be essential to explain some of the seemingly non-local effects observed in flagellar mechanics due to local deformations \cite{Lindemann2005,Pelle2009,Gadelha2013,Coy2017}.
In the model of \cite{Bayly2016}, the cilium is represented by two coupled microfilaments subject to a distribution of internal forces; the interesting result is that, even when the  internal forces are balanced such that they effectively produce no moments on the cilium, oscillations ensue as a result of a flutter instability. 
A derivation of a centerline model based on the two-filament model of \cite{Bayly2016} (results in preparation) would allow rigorous comparisons with other centerline models (\cite{Sartori2016a,Sartori2016b} for example).
Here, for the sake of simplicity and clarity, we analyze the case of a single filament with distributed \textit{axial} forces of constant magnitude, as a useful, albeit idealized, analog for the more complicated centerline force profiles that arise in flagellar mechanics.

\begin{figure}[!t]
	\centering
	\includegraphics[scale=0.225]{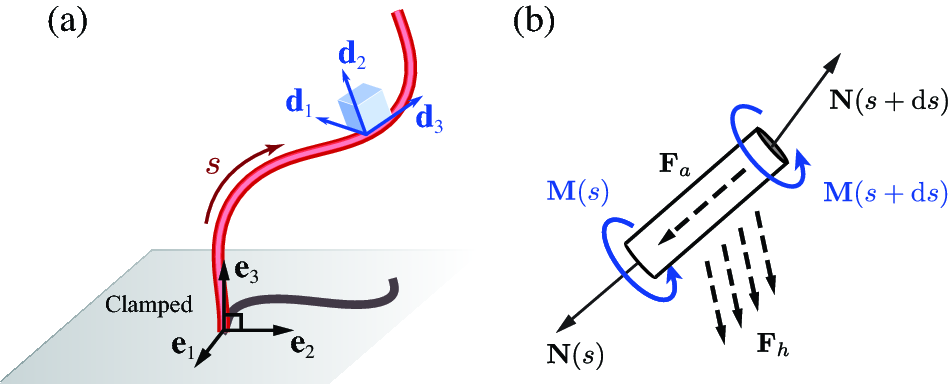}
	\caption{\textbf{The microfilament model.} (a) Schematic of a flexible filament clamped at one end and free at its tip. The filament position is defined by $\mathbf{r}(s, t)$, with $0<s<L$ being the arclength. Its configuration is determined by the orientation of the  material frame $\{\mathbf{d}_1, \mathbf{d}_2, \mathbf{d}_3\}$, with $\mathbf{d}_3 = \partial \mathbf{r}/\partial s$. (b) The forces and moments that apply on an infinitesimal filament element $\mathrm{d}s$. See text for notation.}
	\label{fig:schematic}
\end{figure}

More specifically, we consider an elastic microfilament in a viscous fluid, clamped at one end and acted upon by a distribution of axial forces (see figure~\ref{fig:schematic}). Unlike~\cite{DeCanio2017}, the filament is free to undergo three-dimensional (3D) motions. 
To investigate the filament deformations in 3D, we adapt the numerical framework advanced by \cite{Bergou2008,Gazzola2018},
as well as a combination of linear stability theory and a low-order bead-spring model.
We show that the filament exhibits a Hopf bifurcation that leads to sustained oscillations, consistent with~\cite{DeCanio2017}. 
However, unlike~\cite{DeCanio2017}, the filament undergoes non-planar spinning motions, reminiscent to the cone-like beating motion of cilia, and microtubule streaming motion studied in \cite{Monteith2016}.
Importantly, at larger axial forces, we show the existence of another bifurcation, not reported in~\cite{DeCanio2017}, that causes the filament to transition from 3D spinning to planar wave-like deformations.
We investigate the transition from 3D spinning to 2D flapping under various axial force profiles and in the bead-spring model. We conclude by commenting on the significance of these results to biological cilia and flagella.

\section{Microfilament model}
\label{sec:continuum}

Consider an inextensible elastic filament of length $L$ and circular cross-section of radius $a$,
clamped at its base normally to a fixed plane $(x,y)$; see figure~\ref{fig:schematic}(a). Let 
 $\{\mathbf{e}_1,\mathbf{e}_2,\mathbf{e}_3\}$ be an inertial frame, attached at the filament base, such that $\{\mathbf{e}_1,\mathbf{e}_2\}$ span the $(x,y)$-plane.
The centerline of the filament is described by the vector $\mathbf{r}(s,t)$ where $s$ is arc-length  and $t$ is time. 
The local orientation of the filament is described by an orthonormal material frame $\{\mathbf{d}_1,\mathbf{d}_2,\mathbf{d}_3\}$,
with $\mathbf{d}_3 = {\partial \mathbf{r}}/{\partial s} = \mathbf{t}$ the tangential unit vector. 
The change of  $\{\mathbf{d}_1,\mathbf{d}_2,\mathbf{d}_3\}$ along the filament, at some time $t$,  
is given by
\begin{align}
 \pp{\mathbf{d}_j}{s}  =  \boldsymbol{\kappa} \times \mathbf{d}_j, \quad j = 1,2,3, \label{eq:curv}
\end{align}
where $\boldsymbol{\kappa}$ is a generalized curvature vector recording the infinitesimal change in orientation along the filament; see, e.g.,~\cite{Audoly2010}. 
Specifically, given the rotation matrix $\mathbf{Q}(s,t)$ that maps vectors expressed in the inertial frame to their counterparts in the filament material frame, $\boldsymbol{\kappa}$ is the vector representation, in the material frame, of the skew-symmetric matrix $\boldsymbol{\kappa}^\times = \mathbf{Q}(\mathbf{Q}')^\mathsf{\! T}$, where the prime denotes differentiation with respect to $s$
and $()^\mathsf{\! T}$ is the transpose operator.

\begin{table}
	\begin{center}
		\caption{\textbf{Characteristic scales of cilia and flagella. }}
		\begin{tabular}{lll}
			\textbf{Parameter} & \textbf{Symbol} &  \textbf{Dimensional value} \\[3pt]
			\hline\\[-10pt]
			Bending rigidity &$B$ & $800~\mathrm{pN}\cdot\mu\mathrm{m}^2$ \cite{Xu2016}\\
			Filament length & $L$ & $20~\mu\mathrm{m}$\\
			Fluid viscosity &$\mu\approx\mu_{\text{water},20^\circ}$ & $10^{-3}~\mathrm{Pa}\cdot\mathrm{s}$ \\[3pt]
			\hline\\[-10pt]
			Time scale & $L^4\zeta_\perp/B$ & $0.546~\mathrm{s}$\\
			Frequency scale & $B/(L^4\zeta_\perp)$\; &  $1.83~\operatorname{Hz}$ \\[3pt]\hline\\[-10pt]
			Force scale & $B/L^2$ & $2~\operatorname{pN}$\\
			Force density\; & $B/L^3$ &  $0.1~\operatorname{pN}\cdot\operatorname{\mu m}^{-1}$ \\[3pt]\hline\\[-10pt]
			Slenderness ratio & $a/L$ & $1/100$\\[-3pt]
			Normal drag & $\zeta_{\perp}=\dfrac{4\pi\mu}{\log(L/a)}$ & $2.73\times 10^{-3}~\mathrm{Pa}\cdot\mathrm{s}$
			\\[9pt]
			\hline
		\end{tabular}
		\label{tab:dim}
	\end{center}
\end{table}

The balance of forces and moments on a cross section of the filament are given by the Kirchhoff equations~\cite{Kirchhoff1859}, subject to the clamped-free boundary conditions, 
\begin{align}
 \pp{\mathbf{N}}{s} + \mathbf{f}_h + \mathbf{f}_a = 0, \qquad
 \pp{\mathbf{M}}{s} + \mathbf{t}\times \mathbf{N}  = 0.\label{eq:eom}
\end{align}
Here, $\mathbf{N}$ and  $\mathbf{M}$ are the internal elastic force and bending moment, respectively,  $\mathbf{f}_h$ the hydrodynamic force per unit length, and $\mathbf{f}_a$ the applied force per unit length. 
Taking the cross-product of the tangent vector $\mathbf{t}$ with the balance of moments in~\eqref{eq:eom}, 
the internal elastic force can be readily rewritten as 
\begin{equation}
\mathbf{N}=\mathbf{t}\times({\partial\mathbf{M}}/{\partial s})+\Lambda\mathbf{t}, 
\label{eq:N}
\end{equation}
where $\Lambda=\mathbf{N}\cdot\mathbf{t}$ is a Lagrange multiplier that enforces the inextensiblity constraint $\mathbf{t}\cdot\mathbf{t}=1$. Physically, $\Lambda$ represents the axial tension along the filament.

\begin{figure*}
	\centering
	\includegraphics[scale=0.225]{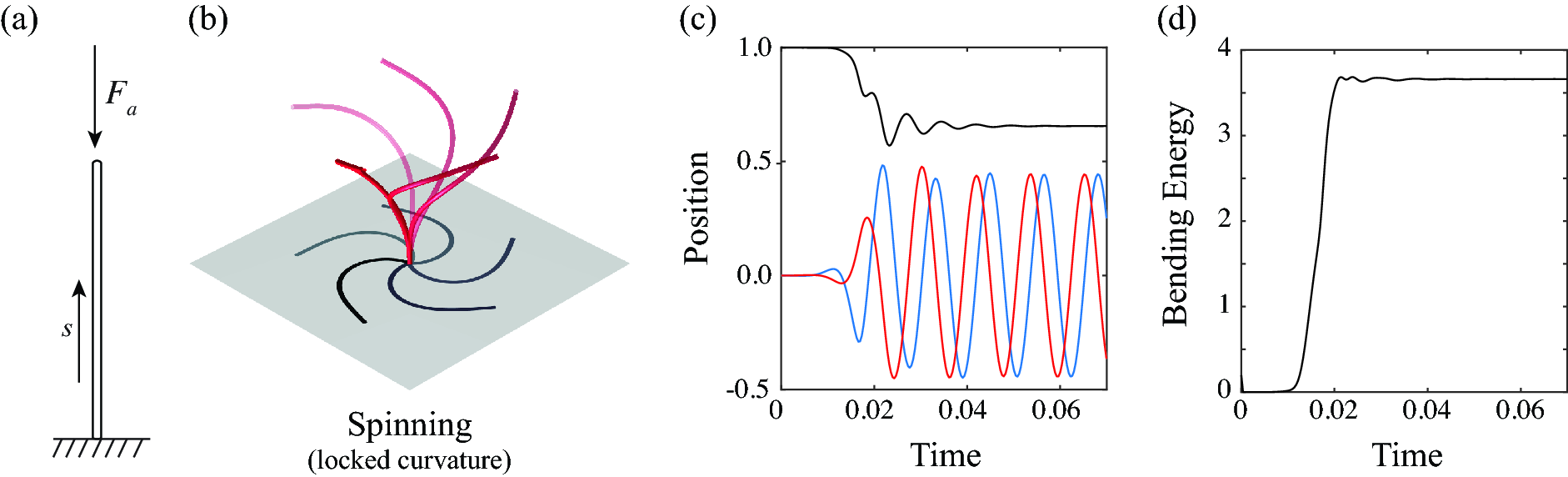}
	\caption{\textbf{Three-dimensional buckling.} (a) Filament subject to an axial follower force of $F_a$, concentrated at its tip. (b) Starting from a straight configuration, the filament buckles into a locked-curvature configuration and undergoes 3D spinning motion for $F_a=75$. (c) The $x$, $y$, and $z$ positions of the filament distal tip point and (d) its bending energy versus time.}
	\label{fig:3Dexample}
\end{figure*}

We take advantage of the small aspect ratio of the filament ($a/L \ll 1$) and use the resistive-force theory that linearly relates $\mathbf{f}_h$  to the instantaneous distribution of velocities 
$\mathbf{v} = \dot{\mathbf{r}}$ 
along the filament centerline, where the dot denotes differentiation with respect to time $t$. 
Specifically, $\mathbf{f}_h= -  \big[\zeta_{\parallel} \vect{t}\mathbf{t}+\zeta_{\perp}(\mathbf{d}_1\mathbf{d}_1 + \mathbf{d}_2\mathbf{d}_2 )\big] \mathbf{v}$ is related to  $\mathbf{v}$
via the anisotropic drag coefficients $\zeta_\perp={4\pi\mu}/{\log(L/a)}$ and $\zeta_\parallel=\zeta_\perp/\gamma$, with $\gamma = 2$,  
(see, e.g.,~\cite{Lauga2009}), equivalently,
\begin{equation}
\begin{split}
\mathbf{f}_h = - {\zeta_{\perp}} \big[\mathbf{I} - \dfrac{\gamma -1}{\gamma} \vect{t}\mathbf{t}\big] \mathbf{v}.
\label{eq:Fh}
\end{split}
\end{equation}
To arrive at~\eqref{eq:Fh},  we introduced the identity matrix  $\mathbf{I}=\mathbf{d}_1\mathbf{d}_1 +\mathbf{d}_2\mathbf{d}_2 + \mathbf{t}\mathbf{t}$.

The force per unit length $\mathbf{f}_a$ emulates the effect of the dynein activity, which we assume to induce a compressive force density $\mathbf{f}_a =  -f_{a} \mathbf{t}$ along the centerline of the filament. This model is reminiscent to the models presented in~\cite{DeCanio2017}, with the distinction that the distribution of the force density may vary along the filament,  $f_a = f_a(s)$, such that the total compressive force is given by $F_a = \int_0^L f_a \mathrm{d}s$.
Non-tangential forces as well as bending and twist moments are ignored. 

We consider a linear constitutive relation between the internal elastic moment $\mathbf{M}$ and $\boldsymbol{\kappa}$. For an axisymmetric filament with circular cross-section of area $A$ and Young's modulus $E$, this linear relation simplifies to $\mathbf{M}=  \mathbf{B} \boldsymbol{\kappa}$, where $\mathbf{B}=B\operatorname{diag}(1,1,2)$ is the bending rigidity tensor, expressed in the filament material frame,  and $B=EA^2$ is the stiffness coefficient. 
It is worth noting that the energy needed to produce twist is of order $L/a$ larger than that to produce bending, therefore we not only  ignore the twist moments due to dynein activity, but also the tangential component of the internal moment, that is, we set $\mathbf{M}\cdot \mathbf{t}=0$, which yields no twist of the filament; see~\cite{Eloy2012}.

In the absence of twist, the generalized curvature $\boldsymbol{\kappa}$ can be expressed as $\boldsymbol{\kappa} = k_1\mathbf{d}_1+k_2\mathbf{d}_2$. Here, $k_1(s,t)$ and $k_2(s,t)$ are scalar curvature functions, related to the  curvature $\kappa(s,t)$ and torsion $\tau(s,t)$ in the Frenet-Serret formulation, such that $\kappa=\sqrt{k_1^2+k_2^2}$ and $\tau={\partial}[\tan^{-1}({k_2}/{k_1)}]/{\partial s}$. It is important to distinguish here between twist and torsion. It is also worth noting that the Frenet-Serret frame $(\mathbf{n},\mathbf{b},\mathbf{t})$, where $\mathbf{n}$ and $\mathbf{b}$ are the normal and binormal unit vectors, is not a material frame. In particular, $\{\mathbf{d}_1,\mathbf{d}_2,\mathbf{d}_3\equiv \mathbf{t}\}$ is related to $(\mathbf{n},\mathbf{b},\mathbf{t})$ through a rigid rotation about the $\mathbf{t}$-direction by an angle equal to $\tan^{-1}({k_2}/{k_1)}$.

The dimensional parameters and characteristic scales relevant for cilia and flagella are summarized in Table~\ref{tab:dim}. Here, we rewrite the system of equations in non-dimensional form using the filament length $L$ as the characteristic length scale and $T = L^4\zeta_\perp/B$ as the characteristic time scale. 
The intrinsic time scale $T$ arises from balancing the elastic and hydrodynamic forces. 
The model admits a second time scale $T_a = \xi L^2/F_a$ that arises from balancing the total applied force with the hydrodynamic force.
The ratio of the two time-scales $T/T_a = F_a L^2/B$ is equivalent to a dimensionless force, which we use in \S\ref{sec:tipforce} as the main parameter to study the filament behavior. 
The non-dimensional system of equations is characterized by the drag anisotropy parameter $\gamma = {\zeta_\perp}/{\zeta_\parallel}=2$, and the parameters that describe the distribution of axial forces along the filament centerline, giving rise to the non-dimensional force $F_a L^2/B$. 
In the following, we consider all variables and parameters to be non-dimensional.

\section{Numerical method}
\label{sec:num}

To solve for the filament dynamics in three-dimensions, we numerically integrate the governing equations~\eqref{eq:eom}, together with the corresponding boundary conditions. To this end, we discretize the filament's centerline into $n+1$ vertices $\mathbf{r}_1, \ldots, \mathbf{r}_{n+1}$ and $n$ straight edges $\boldsymbol{\ell}_i =  \mathbf{r}_{i+1} - \mathbf{r}_i$, where $i = 1,\ldots, n$, in the spirit of \cite{Bergou2008} and \cite{Gazzola2018}. 
The  unit tangent  to edge $i$ is defined as $\mathbf{t}_i =\boldsymbol{\ell}_i / \ell_i$ where $\ell_i = \|\boldsymbol{\ell}_i \|$. 
The translational velocities $\mathbf{v}_1, \ldots \mathbf{v}_{n+1}$ are assigned to vertices.  
The body frames $\{\mathbf{d}_{1,i},\mathbf{d}_{2,i},\mathbf{d}_{3,i}=\mathbf{t}_i\}$ and the rotation matrices $\mathbf{Q}_i$  are naturally assigned to edges.

To obtain a discrete representation of the generalize curvature vector $\boldsymbol{\kappa}$, we start from
the definition of the associated skew-symmetric matrix $\boldsymbol{\kappa}^\times = \mathbf{Q}(\mathbf{Q}')^\mathsf{T}$. The solution to this first-order differential equation is of the form $\mathbf{Q}(s+\Delta s) = \exp(-\boldsymbol{\kappa}^\times  \Delta s)\mathbf{Q}(s)$, where $\Delta s$ is a segment of constant curvature.  Thus, writing  $\mathbf{Q}_i = \exp(-\boldsymbol{\kappa}_{i}^\times \overline{\ell}_i)\mathbf{Q}_{i-1}$, we define the discrete curvature matrix as  $\boldsymbol{\kappa}_{i}^\times=-{\log\big(\mathbf{Q}_{i}\mathbf{Q}^\mathsf{T}_{i-1}\big)}/{\overline{\ell}_i}$, 
where $\overline{\ell}_i=\f12 (\ell_{i-1}+\ell_{i})$ is a `Voronoi' integration domain from the midpoint of the previous edge to that of the next edge.
Given~$\boldsymbol{\kappa}_{i}^\times$, we can readily evaluate $\boldsymbol{\kappa}_{i}$ and the discrete elastic moments $\mathbf{M}_i$ in the filament  material frame.

We solve for the discrete elastic forces $\mathbf{N}_{i}$  in the material frame using~\eqref{eq:N}. The inextensibility constraint is enforced weakly by considering a large tensile stiffness ${E}A$ along the filament's centerline. 
Next, we transform $\mathbf{N}_i$ to the inertial frame, and substitute into the balance of forces in~\eqref{eq:eom} using the expression for the hydrodynamic force density $\mathbf{f}_h$ from \eqref{eq:Fh} to get the inertial frame velocities $\mathbf{v}_i$.
We use standard time integrators for stiff equations (MATLAB's \verb|ode15s|) to propagate the filament position $\mathbf{r}_i$ forward in time. To close the system, we enforce the clamped boundary condition by fixing $\mathbf{r}_1=\mathbf{0}$ and $\mathbf{t}_1=\mathbf{e}_3$, while leaving the free end unconstrained.  Additional details on the discretization method and numerical validation are included in the supplemental document.

\begin{figure*}[!htb]
	\centering
	\includegraphics[scale=0.225]{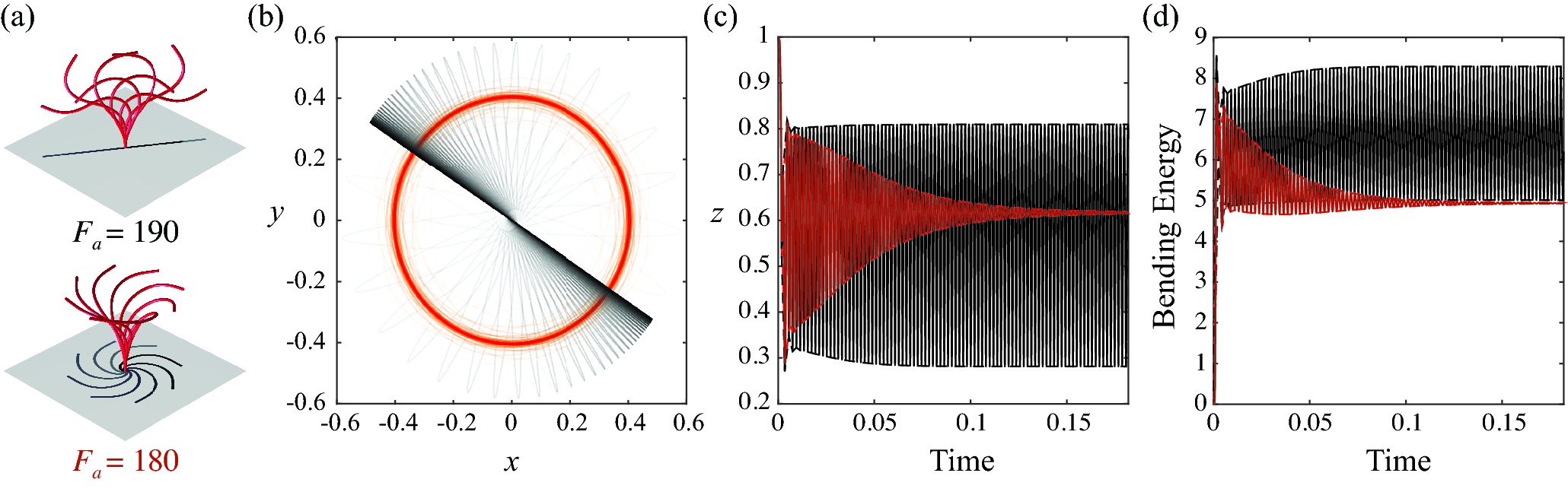}
	\caption{\textbf{Transition from 3D spinning to 2D flapping.} (a)  3D spinning motion at $F_a=180$ versus  2D flapping motion at $F_a=190$. (b) when projected onto the $(x,y)$-plane, the tip of the spinning filament traces a circular trajectory, while the trajectory of the tip of the flapping filament collapses to a fixed line. (c) and (d) shows that both the $z$ position of the distal tip and the total bending energy of the filament become constant in time for spinning, but remain oscillatory for flapping.}
	\label{fig:transition}
\end{figure*}

\section{Linear stability analysis}
\label{sec:lsa}

Consider a straight filament subject to a distribution of axial forces $\mathbf{f}_a = -f_a(s)\mathbf{t}$, along the filament centerline, where $f_a(s)$ is a general function of $s$.  In order to assess the stability of the straight configuration, we linearize the equations of motion about the straight equilibrium and solve for the dominant eigenvalue.

To begin, we substitute~\eqref{eq:N} and~\eqref{eq:Fh} into the force balance in~\eqref{eq:eom} to arrive at the vector equation 
\begin{equation}
\dot{\mathbf{r}} = \gamma (\Lambda' + \kappa \kappa' - f_a)\mathbf{t} + (\Lambda k_2 - k_2'')\mathbf{d}_1 - (\Lambda k_1 - k_1'')\mathbf{d}_2,
\label{eq:rdot}
\end{equation}
where we used the fact that $\kappa = \sqrt{k_1^2 + k_2^2}$. 
Equation~\eqref{eq:rdot} leads, upon further simplifications (see details in the supplemental document),  to three nonlinear scalar equations in terms of the filament tension $\Lambda(s,t)$ and filament shape: curvature $\kappa(s,t)$ and torsion $\tau(s,t)$. Expanding these equations about the straight equilibrium state $\kappa =\tau =0$, and linearizing with respect to $\kappa$ and $\tau$, it can be shown that above vector equation leads to two scalar equations only, 
\begin{align}
\Lambda''=&\ f_{a}' \label{eq:constraint}\\
\kappa'''' - \Lambda \kappa'' + \gamma \dot{\kappa}=&\ 0.
\label{eq:linDyn}
\end{align}
The third equation is trivially satisfied at the linear level.
Since these equations do not depend on torsion $\tau(s,t)$, the linear analysis cannot capture 3D deformations of the filament. That is to say, in the linear regime near the straight equilibrium, the filament undergoes \textit{planar deformations}. Thus, without loss of generality, we can equate the arc-length $s$ with the $z$ coordinate in the inertial frame and set $\kappa = y''$, where the prime notation here denotes differentiation with respect to $z$. Substituting into~\eqref{eq:linDyn} and integrating  with respect to $z$ twice, we arrive at the fourth-order partial differential equation,
\begin{equation}
y'''' - \Lambda y'' + \gamma \dot{y} = 0,
\end{equation}
for the infinitesimal displacement $y(z,t)$ in the $(y,z)$-plane. 
To close this equation, we integrate \eqref{eq:constraint} from the free tip of the filament and define the internal traction force
\begin{equation}
 \Lambda(z) =\int_L^z f_a(\tilde{z})\,\dif \tilde{z}. 
\end{equation}
Assume the solution $y(z,t)$ is separable in the form of $y(z,t)=Y(z)e^{\lambda t}$, we obtain the boundary value problem
\begin{equation}
\begin{split}
Y''''- \Lambda Y''+ \lambda \gamma Y=0,\\
Y(0)=Y'(0)=0,\quad Y''(L)=Y'''(L)=0.  \label{eq:BVP}
\end{split}
\end{equation}
The first two boundary conditions correspond to the clamped end and the latter two to the free end.
In general, the differential equation in~\eqref{eq:BVP} belongs to the class of linear differential equations with non-constant coefficients, given that $\Lambda(z)$ is a function of $z$ for a general distribution of axial forces. Therefore, depending on the form of $f_a(z)$, this boundary value problem may not admit an analytical solution.
The boundary conditions welcome the use of the free vibration modes of a clamped-free beam as test functions for a numerical solution of this boundary value problem.
Namely, we expand $Y(z)$ onto the eigenfunctions of the clamped-free linear beam equations
\begin{equation}
Y(z) = \sum_{i=1}^{\infty}Y_i\Phi_i(z),
\end{equation}
where $Y_i$ are constant coefficients and $\Phi_i(z)$ are given by
\begin{equation}
\begin{split}
\Phi_i(z) &= \sum_{n=0}^{\infty}\cosh(a_n z) - \cos(a_n z)  \\
&\hspace{0.5in} - b_n[\sinh(a_n z) - \sin(a_n z)],
\end{split}
\end{equation}
with $b_n = (\cos a_n + \cosh a_n)/(\sin a_n + \sinh a_n)$ and $a_n$ the $n$-th root of the transcendental characteristic equation $\cos a_n \cosh a_n+1=0$. 
{This ``assumed-modes'' method, an extension of the classical Ritz method \cite{Ritz1909,Macdonald1933}, was used to analyze the stability of flagella dynamics in \cite{Bayly2015,Bayly2016}.}

Finally, we project~\eqref{eq:BVP} onto these basis functions and rewrite it in  weak form as
\begin{align}	
Y_i \left[ \int_{0}^{L}\Phi''_i\Phi''_j\,\dif z \right. & - \int_{0}^{L}   \Phi_i\Lambda\Phi''_j\,\dif z \nonumber \\[-1ex] 
& \left. +  \ \lambda {\gamma}\int_{0}^{L}\Phi_i\Phi_j\,\dif z \right]=0.
\label{eq:wBVP}
\end{align}
This equation can be rewritten in matrix form as $(\mathbf{K} + \mathbf{C} + \lambda \gamma \mathbf{M})\mathbf{Y}=0$. Thus, the linear stability of the filament away from the straight configuration can be obtained via the eigenvalue problem $\mathbf{K} + \mathbf{C} = -\lambda {\gamma}\mathbf{M}$.
The real part of the resulting $\lambda$ indicates the growth (or decay) of pertubations and the imaginary part correspond to the frequency of a given modal shape ${Y}_i\Phi_i(z)$.

\section{Spinning versus flapping oscillations}
\label{sec:tipforce}

We  first consider the set-up analyzed in~\cite{DeCanio2017} of an elastic filament subject to a follower axial force  of magnitude $F_a$ concentrated at the distal tip of the filament; see figure~\ref{fig:3Dexample}(a). 
In~\cite{DeCanio2017}, the filament is confined to undergo planar motions. Starting from a small perturbation away from the straight configuration, the authors identify three dynamical behaviors depending on the magnitude of the axial force $F_a$: 
(i) for small force, the filament returns monotonically to its original straight configuration, 
(ii) as the force value increases, the filament displays decaying oscillations back to the straight configuration, 
and (iii) for forces above a given threshold, the filament settles into finite-amplitude periodic deformations, which they referred to as \textit{planar flapping}.

Here, we start with a small \textit{non-planar} perturbation away from the straight configuration and we numerically solve for the filament dynamics. 
In figure~\ref{fig:3Dexample}, we set $F_a = 75$, which lies in the planar flapping regime of~\cite{DeCanio2017}. 
 Unlike the flapping behavior reported in the latter, the filament buckles into a configuration with locked curvature and undergoes three-dimensional spinning about the $z$-direction as illustrated in figure~\ref{fig:3Dexample}(b). The tip displacement of the filament and its bending energy as a function of time are reported in figure~\ref{fig:3Dexample}(c) and (d), respectively. The $(x,y)$ coordinates of the filament tip settle into periodic motions while the $z$-coordinate goes to a constant value. The filament bending energy also goes to a constant value, emphasizing that, in its buckled state, the curvature of the filament is constant (see supplemental movie S1).

\begin{figure}[!t]
	\centering
	\includegraphics[scale=0.225]{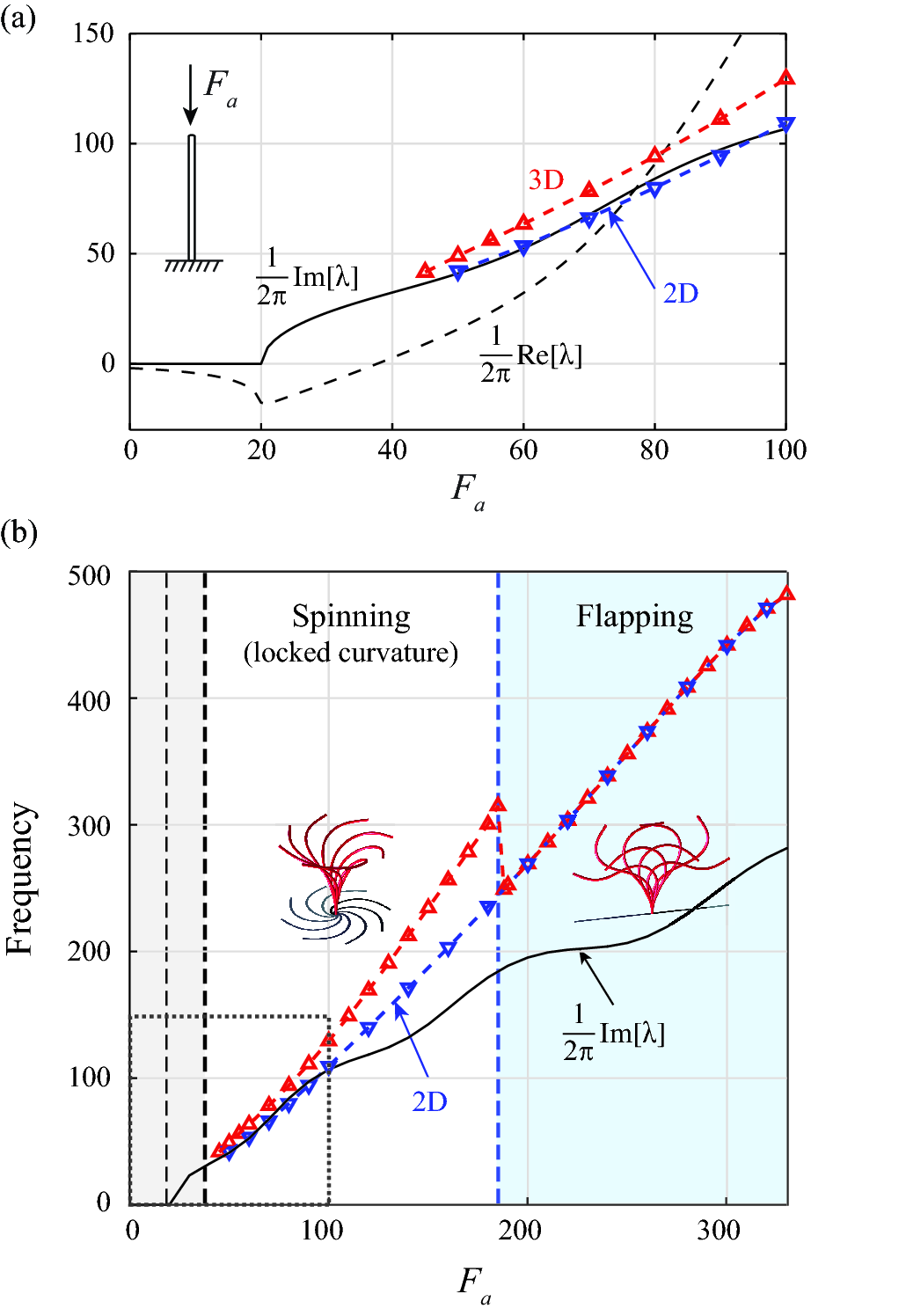}
	\caption{\textbf{Frequency of oscillations versus linear stability analysis.} (a) real (dashed line) and imaginary (solid black line) parts of the dominant $\lambda$ as a function of $F_a$ for the range of $F_a$ reported in~\cite{DeCanio2017}.  A Hopf bifurcation at $F_a \approx 37.7$ marks the transition from a stable straight configuration to sustained oscillations. The nonlinear oscillation frequencies are computed for \textit{non-planar} perturbations (red line) and confined \textit{planar} perturbations (blue line). Confined filaments oscillate at frequencies that are in reasonable agreement with those predicted by the linear stability analysis, while 3D spinning motions occur at higher frequencies. The relative frequency difference ($1-$flapping/spinning frequency) ranges from $\sim 15\%$ to $\sim 20\%$. (b) Same analysis extended to a larger range of $F_a$ -- dashed box in lower left corner corresponds to the results in (a). We observe a second transition at $F_a \approx 188$ from stable 3D spinning to stable 2D flapping motions, even when the filament is subjected to \textit{non-planar} perturbations.  At large values of $F_a$, the flapping frequency deviates significantly from the linear stability analysis.
	}
	\label{fig:transitionF}
\end{figure}

By symmetry of the geometry and material properties of the filament, planar initial perturbations lead to filament motions that remain confined to the perturbation plane.
In particular, for planar initial perturbations, the filament in~figure~\ref{fig:3Dexample}  undergoes 2D flapping motions as reported in~\cite{DeCanio2017} (see supplemental movie S2). However, non-planar perturbations lead to 3D spinning motions (see figure~\ref{fig:3Dexample} and supplemental movie S1). 
Taken together, these results go beyond the findings of~\cite{DeCanio2017} to show that,  
subject to non-planar perturbations, planar flapping can also be unstable.

We next examine the filament behavior for increasing values of $F_a$. We observe a transition from 3D spinning  to 2D wave-like oscillations as shown in figure~\ref{fig:transition}(a). For $F_a = 180$, the filament settles into a buckled configuration with locked curvature and undergoes 3D spinning about the $z$-direction, whereas for $F_a = 190$, the filament converges in finite time to planar oscillations. 
The $(x,y)$ motion of the filament tip is shown in figure~\ref{fig:transition}(b): the red line corresponds to 3D spinning ($F_a = 180$) and the black line to 2D flapping ($F_a = 190$). The flapping motion occurs in an arbitrary plane, that depends on the non-planar initial perturbations. However, the frequency and amplitude of these wave-like deformations are independent of initial conditions. For  3D spinning, the $z$-coordinate of the filament tip and the filament bending energy go to constant values in finite time; see red lines  in figure~\ref{fig:transition}(c) and (d), respectively. For the 2D flapping motion, they oscillate with constant amplitudes (black lines).  It is important to distinguish between the 2D flapping motion obtained here and the planar motions of~\cite{DeCanio2017}. The planar motions in~\cite{DeCanio2017} are unstable to non-planar perturbations and are obtained as a result of confining the filament in one plane. Here, the filament dynamically converges to planar deformations, even when subject to large non-planar initial perturbations (see supplemental movie S3). In other words, for this value of $F_a$, the planar motion is robust to non-planar initial perturbations.

We plot in figure~\ref{fig:transitionF}(a) the oscillation frequency of the filament as a function of $F_a$ for the range of values considered in~\cite{DeCanio2017}.
We compare these frequencies to the real and imaginary values of the dominant eigenvalues obtained from the linear stability analysis about the straight configuration discussed in~\S\ref{sec:lsa}. 
Linear stability analysis shows that the imaginary part of the dominant eigenvalue (solid black line) becomes complex at $F_a \approx 20$ whereas the real part (dashed black line), associated with the rate of growth of the initial perturbation, remains negative until $F_a\approx 37.7$.    That is to say, the straight filament configuration is monotonically stable for $F_a < 20$ and
stable with decaying oscillations for $20< F_a < 37.7$. The filament undergoes a Hopf bifurcation at $F_a = 37.7$, as the complex pair of eigenvalues crosses the imaginary axis, leading to unstable motions with growing oscillations. The linear oscillation frequencies are given by the imaginary part of the dominant eigenvalue Im[$\lambda$]/$2\pi$. 
The linear stability results are quantitatively consistent with those in~\cite{DeCanio2017}.

\begin{figure}[!t]
	\centering
	\includegraphics[scale=0.225]{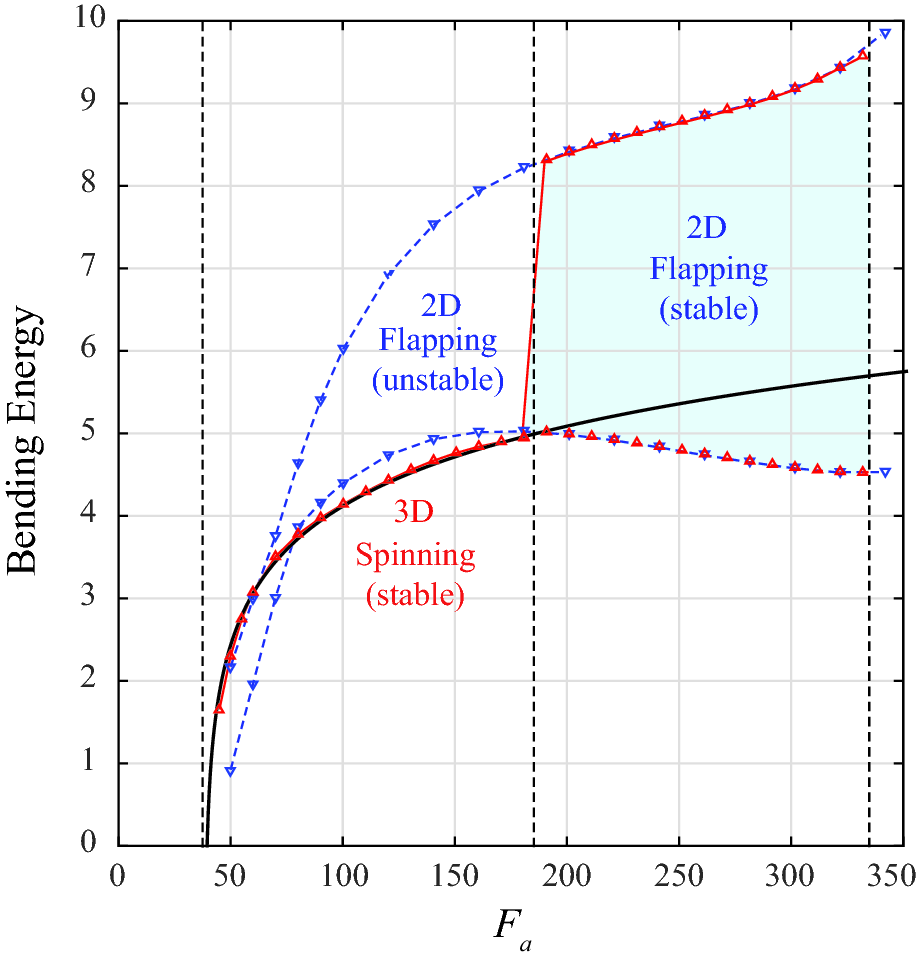}
	\caption{\textbf{Bending energy versus axial force.} The bending energy of spinning motion (red) increases logarithmically with $F_a$ following the curve $E_{\rm bending} = \ln(F_a-38.7)$ (solid black line). The envelope of the flapping bending energy, from minimum to maximum energy,  is depicted for the unstable (confined) 2D flapping for $F_a < 188$ (blue) and the stable 2D flapping for $F_a > 188$ (red and blue).  The minimum of the flapping bending energy of flapping starts to decrease near the transition.}
	\label{fig:BE}
\end{figure}

The frequency of the nonlinear oscillations is computed by looking at correlations of peaks in the time evolution of the average position of the filament. 
Nonlinear oscillations are analyzed for non-planar and planar (confined) perturbations, depicted in red and blue, respectively. For $F_a > 37.7$, non-planar perturbations give rise to 3D spinning motions at frequencies that are consistently larger than the frequencies of the 2D deformations associated with planar perturbations.
The latter are comparable to the frequencies predicated by the linear stability analysis.
Although the linear analysis correctly captures the transition from stable to unstable straight configuration, it only predicts planar deformations and cannot distinguish between 2D and 3D oscillations. The fact that the linear analysis only captures planar deformations is due to the structure of the linear equations, which decouples torsion from bending curvature, as discussed in~\S\ref{sec:lsa}.  Our nonlinear analysis indicates that, subject to non-planar perturbations, the Hopf bifurcation at $F_a = 37.7$ leads to a buckled configuration with locked curvature that spins at a higher frequency (around $ 15\%$ larger) than the frequency of confined deformations (see figure~\ref{fig:transitionF}a). 

In figure~\ref{fig:transitionF}(b), we extend this analysis to larger values of $F_a$, with the lower left corner, highlighted by a narrow dashed line corresponding to  figure~\ref{fig:transitionF}(a).   Our analysis indicates the presence of another bifurcation at  $F_a \approx 188$ that causes the filament to transition from 3D spinning to 2D flapping motions, even when subjected to non-planar perturbations. At these large values of $F_a$, the flapping frequency deviates significantly from the linear stability analysis; {the latter fails to capture the second transition from 3D spinning to 2D flapping.} We emphasize that these flapping motions are fundamentally distinct from the flapping motions reported in~\cite{DeCanio2017} for $F_a < 100$. The latter are unstable to non-planar perturbations. For $F_a > 188$, the filament dynamically converges to planar flapping.

{Before we proceed, a few comments on the scaling of the dimensional force and instability threshold with the filament length are in order. The dimensional force is equal to $F_aB/L^2$. Thus, in dimensional form, for a filament of length $L=20$ $\mu$m (see Table~\ref{tab:dim}),  the first transition from straight configuration to 3D spinning occurs at $F_a B/L^2 =37.7B/L^2 = 75.4$ pN and the second transition from 3D spinning to 2D flapping occurs at $F_a B/L^2 =188 B/L^2 = 376$ pN. Shorter filaments require larger values of the dimensional force to trigger these instabilities.}

We examine more closely the robustness of the 3D spinning and 2D flapping motions with respect to the amount of initial perturbation away from the straight equilibrium. 
Our numerical experiments show that, for $37.7<F_a < 188$, any minute perturbation to the planar symmetry puts the filament into a 3D spinning state, whereas $F_a> 188$ always leads to planar flapping even for dramatically non-planar initial conditions. These results indicate that the second transition is not sensitive to initial perturbations.  They further suggest  
that for $37.7<F_a < 188$, the basin of attraction of the 3D spinning mode is the set of all non-planar initial conditions, while for $F_a> 188$, the basin of attraction of the 2D flapping is the full space of initial conditions, planar and non-planar (See supplemental movies S3 and S4).

To elucidate the physical mechanisms at play in these regimes, we note that the work done by the axial force is balanced by both the elastic bending energy of the filament and the work dissipated via viscous drag due to the filament motion.
In 3D spinning regime, the filament maintains a locked curvature, characterized by a constant bending energy. In figure~\ref{fig:BE}, we plot the bending energy of the filament versus the force value $F_a$ for both 3D spinning (red) and planar flapping (blue) which are unstable for $F_a < 188$. 
In its locked-curvature configuration, the bending energy grows  logarithmically with $F_a$.
When undergoing 2D flapping deformations, the bending energy accesses larger ranges of energy values. 
This explains why the spinning frequencies are larger than the frequencies of confined 2D flapping: since a smaller amount of the work done by the axial force is absorbed by the bending energy of the locked filament than by the confined 2D flapping filament, more work is available for 3D spinning. 
In the 2D flapping regime, the work done by the axial force is too large; the filament cannot settle on a locked-curvature configuration that stores a sufficient amount of bending energy to allow the excess work to be dissipated via viscous drag by a rigid rotation of the locked filament.  The filament has to deform.

To summarize, the results in this section show that the nonlinear dynamics of an elastic filament in a viscous fluid, clamped at one end and subject to a follower compressive force at its free end, is far richer than previously recognized.  
As the magnitude of the axial force $F_a$ increases, four distinct regimes of dynamical behaviors are observed:  (i) a monotonically stable regime, (ii) a regime with decaying oscillations, (iii)  3D spinning motions at a buckled configuration with locked curvature, 
and (iv)  2D flapping motions where the filament dynamically converge to  planar oscillations.

\begin{figure}[!t]
	\centering
	\includegraphics[scale=0.225]{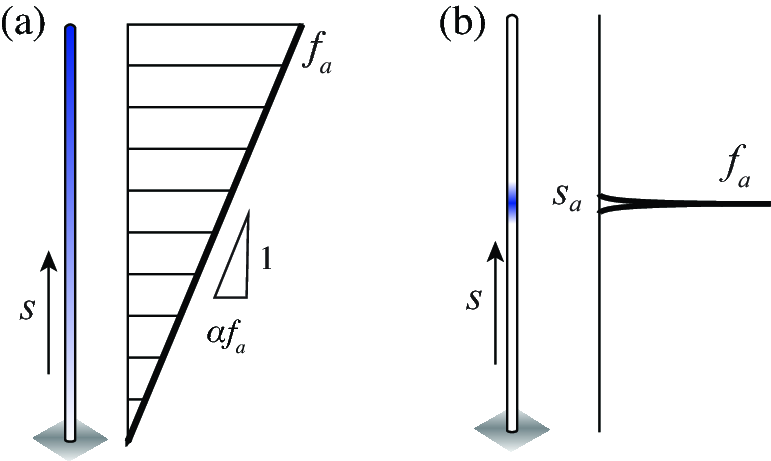}
	\caption{\textbf{Axial forces.} Schematic showing the distribution of the axial forces along the filament centerline: (a) an axial force density that varies linearly from the base of the filament to its tip, and (b) the axial force is concentrated at a distance of $s_a$ from the filament base.}
	\label{fig:axialforce}
\end{figure}

\section{Axial force profiles}
\label{sec:gallery}

\begin{figure*}[!t]
	\centering
	\includegraphics[scale=0.225]{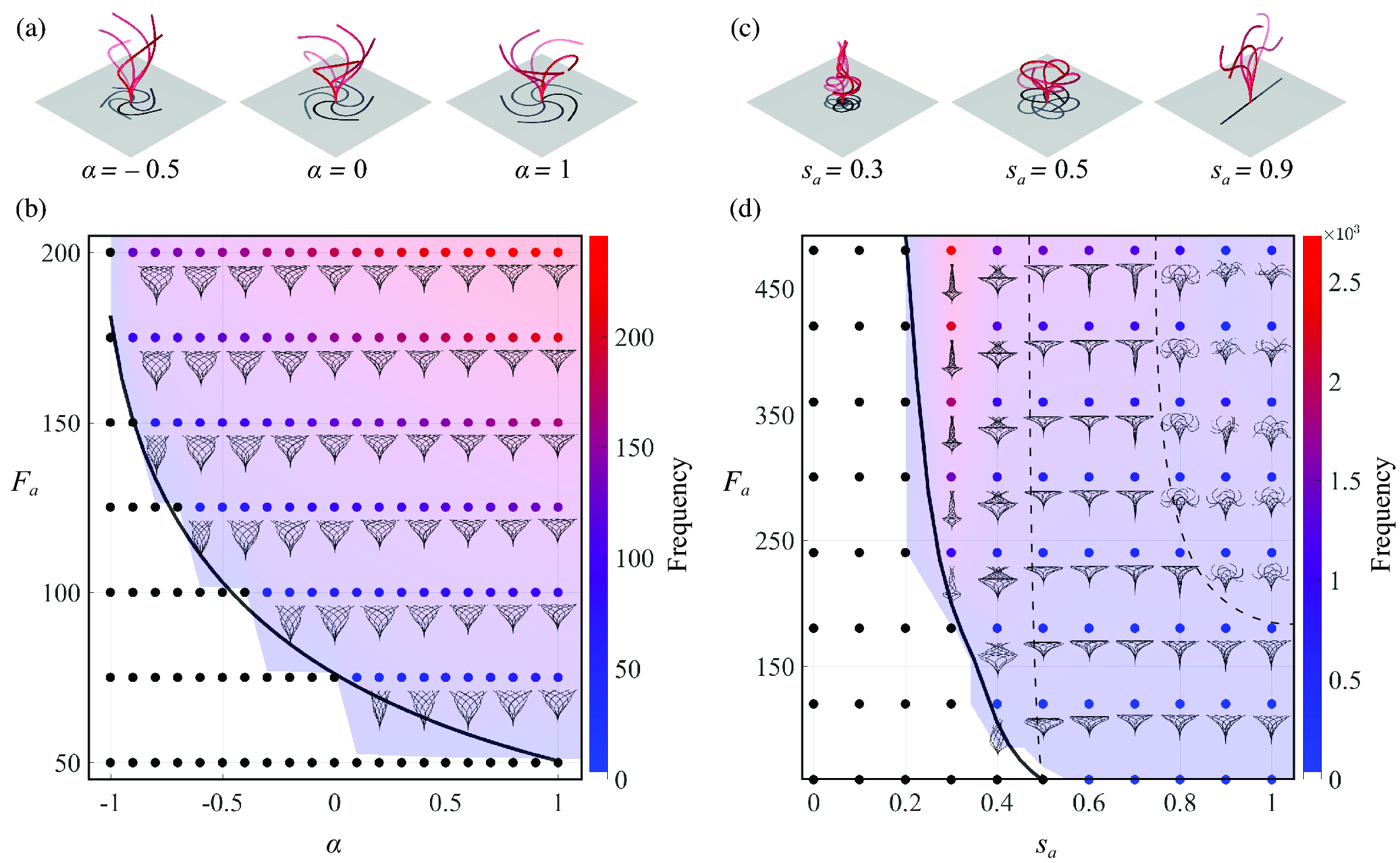}
	\caption{\textbf{Gallery of filament oscillations.} In (a) and (b), filaments subject to linear distribution of axial forces exhibit buckling and 3D spinning motion only despite changes in parameter $\alpha$. But the threshold of instability decreases and the radius of the circle formed by the filament tips increases for higher $\alpha$. In (c) and (d), filaments are subjected to a concentrated axial force at varying location $s_a$ sees qualitatively distinct regimes of motion. For $s_a<0.5$, the distal tip of the filament remain taller than the largest circle formed by the spinning motion.  When forces are concentrated closer to the tip, we observe the 3D spinning to 2D flapping transition of figure~\ref{fig:transition}. Note that in both (b) and (d), changes in the threshold of initial instability of static straight equilibrium are well-matched by the linear stability analysis (solid black lines). $F_a=125$ and $300$ respectively for snapshots shown in (a) and (c).}
	\label{fig:gallery}
\end{figure*}

We examine the robustness of the four regimes identified in~\S\ref{sec:tipforce} to the distribution of axial forces along the filament centerline.
Specifically, we consider two force profiles. The first profile consists of a linear force density, with two parameters $\alpha$ and $f_a$,
\begin{align}	
\mathbf{f}_a=-f_a\left[1+\alpha\left(2s-1\right)\right] \mathbf{t}. 
\label{eq:fL}
\end{align}
Here, $f_a$ is the intensity of the force density, $\alpha$ ranges from $-1$ to $1$ and $\alpha f_a$ determines the slope of the linear force distribution; negative $\alpha$ corresponds to a force density that decreases as $s$ increases towards the tip of the filament and positive $\alpha$ to a force density that increases towards the filament tip. The total compressive force exerted on the filament is given by $F_a = \int_0^1 f_a \left[1+\alpha\left(2s-1\right)\right]\mathrm{d}s = f_a$. 
The second profile consists of a concentrated force $f_a$ at a distance $s_a$ from the base point of the filament
\begin{align}
\mathbf{f}_a = - f_a\delta(s-s_a) \mathbf{t}, 
\label{eq:fD}
\end{align}
where $\delta(s-s_a)$ is the Dirac delta function and $s_a$ ranges between $0$ to $1$; $s_a =1$ corresponds to the force at the tip of the filament. The total compressive force exerted on the filament is given by $F_a = \int_0^1 f_a \delta(s-s_a)\mathrm{d}s = f_a$. A schematic depicting these two profiles is shown in figure~\ref{fig:axialforce}.

In figure~\ref{fig:gallery}(a), we set $F_a = 125$ and consider three values of $\alpha$.  For $\alpha = -0.5$ the axial force decreases towards the filament tip, $\alpha = 0$ corresponds to a constant force density along the filament, and for $\alpha = 1$ the axial force increases towards the filament tip. In all three cases, the filament buckles into a locked configuration and undergoes 3D spinning motion. We systematically analyze the behavior of the filament as a function of the two parameters $\alpha$ and $F_a$. The results are shown in the parameter space $(\alpha, F_a)$ in figure~\ref{fig:gallery}(b): black dots correspond to cases where the straight configuration is stable, the solid black line denotes the Hopf bifurcation marking the transition to growing oscillations based on the linear stability analysis of \S~\ref{sec:lsa}. The colormap denotes the frequency of the nonlinear oscillations of the filament. The filament motion is represented next to each point $(\alpha, F_a)$ by superimposing the filament configuration at various time steps within the oscillation cycle. 
Force densities that decrease towards the tip of the filament ($\alpha < 0$) require larger force values $F_a$ to reach the threshold for the Hopf bifurcation. 
In all cases past the Hopf bifurcation, the filament buckles and undergoes 3D spinning at frequencies larger than the frequencies predicted by the linear stability analysis. 

In figure~\ref{fig:gallery}(c), we consider a concentrated force $F_a = 300$ and vary its location along the filament. At $s_a = 0.3$, the filament buckles into a locked curvature and spins about the $z$-direction. The filament exhibits large curvatures near the point of application of the axial force, with flatter curvature near the tip. For $s_a = 0.5$, the filament also buckles, exhibiting similar behavior but more gradual variations in curvature. For $s_a=0.9$ the filament undergoes planar flapping motion, even when subject to non-planar perturbations. A gallery of the filament behavior as a function of the two parameters $s_a$ and $F_a$ is depicted in figure~\ref{fig:gallery}(d). The filament  first transitions from a straight configuration to 3D spinning motion. A second transition occurs from 3D spinning to 2D flapping motion as $s_a$ and $F_a$ increase further, as highlighted by the dashed black line in the upper right corner of the parameter space.
In dimensional form, the force threshold $F_a B/L^2$ required to trigger these instabilities increases as the filament length decreases.

\begin{figure*}[!htb]
	\centering
	\includegraphics[scale=0.225]{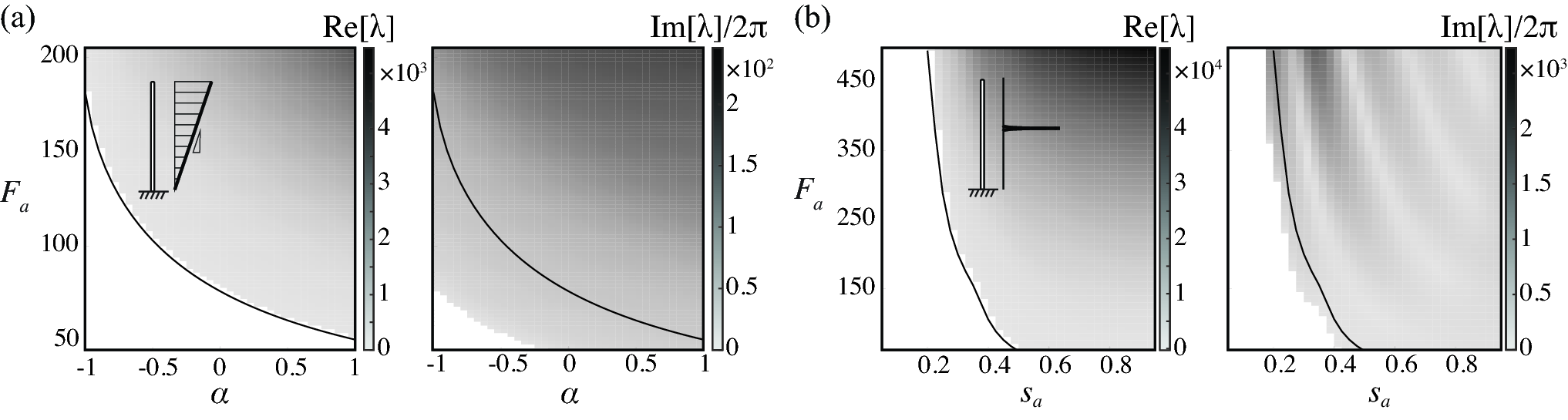}
	\caption{\textbf{Linear stability analysis.} We plot the full results of linear stability analysis for (a) the linear forcing profiles and (b) the concentrated forcing profiles. The real parts of $\lambda$ indicate the stability of straight equilibrium, while the imaginary parts predict the frequency of oscillations. Similar to figure~\ref{fig:transitionF}, the imaginary parts become positive before the real parts, indicating the existence of decaying oscillations. The frequencies predicted here by linear stability are in rough agreement with those shown in figure~\ref{fig:gallery}. }
	\label{fig:gallery_stability}
\end{figure*}

For both profiles of axial forces, the first transition, marking the destabilization of the straight configuration under the distribution of axial forces $F_a$, is captured by the linear stability analysis of~\S\ref{sec:lsa}. As we vary the two parameters corresponding to each forcing profile, $(\alpha, F_a)$  and $(s_a, F_a)$, respectively,  linear stability analysis indicates a transition from stable to unstable straight configuration through a Hopf bifurcation as shown in figure~\ref{fig:gallery_stability}. However, linear stability analysis predicts growing planar oscillations at frequencies less than the frequencies of the 3D spinning motions obtained in the nonlinear system. Further, this linear analysis does not capture the second transition, from non-planar spinning to planar flapping, observed in the concentrated force profile.

Based on the gallery of filament oscillations reported in figure~\ref{fig:gallery}, the following observations are in order: (1) The first buckling instability, captured by the linear stability stability analysis, always leads to 3D spinning motions. (2) The second transition from 3D spinning to 2D flapping seems to occur when the force distribution is biased towards the tip of the filament; we explored this observation in more detail next. (3) Away from the transition thresholds, the filament behavior is insensitive to changes in the parameters of the axial force, even to changes in the form of the axial force profile. These observations could have important implications on the relevance of this instability-driven mechanism to biological cilia and flagella as discussed in~\ref{sec:discussion}.

To better understand the effect of the distribution of axial forces along the filament on this transition from 3D spinning to 2D flapping, we introduce a forcing profile that allows us to continuously change from a linearly-distributed force profile to a force concentrated at the filament tip. To this end, we consider
\begin{equation}
\mathbf{f}_a = - f_a (p+1){s}^p\,\mathbf{t},
\label{eq:fLD}
\end{equation}
where  $p$ controls how much force is concentrated near the tip in a gradual way.
At $p=1$, the force expression in~\eqref{eq:fLD} is identical to that in~\eqref{eq:fL} for $\alpha = 1$. As $p\to \infty$, this force converges to~\eqref{eq:fD} for $s_a = 1$. Again, we have $F_a = \int_0^1 f_a (p+1)s^p\mathrm{d}s = f_a$

\begin{figure}[!t]
	\centering
	\includegraphics[scale=0.225]{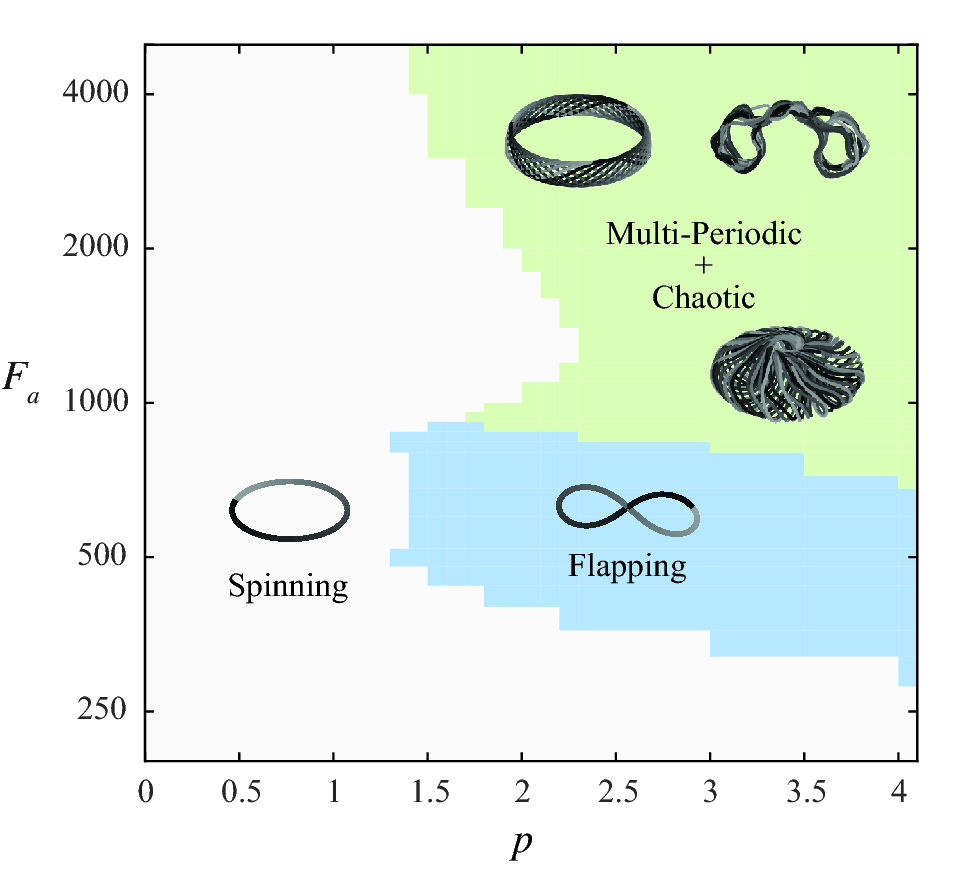}
	\caption{\textbf{Phase diagram based on power-law forcing profiles.} We expand findings of figure~\ref{fig:gallery} by showing the transition behaviors at much higher $F_a$ using the power-law forcing profiles of \eqref{eq:fLD}. We see that only when the axial forces increase at a power larger than $\approx1.5$ would 3D spinning transitions to 2D flapping. Moreover, at very large forces, we can also obtain much more exotic trajectories. The patterns shown above are formed by the average positions of the filament over its arclength (grey scale represents passage of time). }
	\label{fig:powerPhase}
\end{figure}

Figure~\ref{fig:powerPhase} is a phase diagram that summarizes the behavior of the filament as a function of $p$ and $F_a$. We find that for a big range of axial forces, potentially exceeding the biologically-relevant range, when $p=1$, that is, for a linear force distribution, there is no transition from 3D spinning to 2D flapping motions. As $p$ exceeds 1, that is, for force distributions that are larger towards the filament tip, we observe a transition from spinning to flapping motions at biologically-relevant values of the axial force $F_a$ (as discussed in \S\ref{sec:discussion}). For even larger forces, another transition occurs from planar flapping to toroidal flapping motions, and even to chaotic-like behavior. Representative trajectories of the mean position of the filament are superimposed onto the phase space to illustrate the complex nature of these motions.
These results indicate that higher concentrations of axial forces near the distal portion of the microfilament accelerate the development of complex behaviors. 

It is worth noting here that we calculated the Lyapunov exponents for large values of $F_a$ concentrated at the tip of the filament leading to chaotic-like behavior. Namely, we calculated the Lyapunov exponents for arbitrary initial conditions based on \textit{finite separation} in the mean position of the filament (results not shown). Upon the removal of a rotational symmetry due to the initial Euler buckling, trajectories from two nearby initial conditions diverge initially, but saturate at a large distance, giving rise to Lyapunov exponents that are consistent with chaos.
Similar chaotic trajectories are reported in recent models of active biological filaments~\cite{Pearce2018}.
It is not clear if this chaotic behavior is biologically-relevant for a beating flagellum because these large $F_a$ may lie beyond the capabilities of the dynein molecular motors. 
However, it is not completely surprising to obtain chaotic behavior in this dissipative system due to the resemblance of the filament equations of motion to the Kuramoto-Sivashinsky equation that models diffusive instabilities in a laminar flame front.

To conclude this section, we note that the spinning to flapping transition is insensitive to the anisotropy in the fluid drag force. Our numerical experiments show that when changing $\gamma$, the filament continues to transition from 3D spinning to planar flapping then to 3D motion, albeit requiring a different amount of axial force. This observation will allow us to ignore drag anisotropy in our reduced-order, bead-spring model discussed next.

\begin{figure*}[!htb]
	\centering
	\includegraphics[scale=0.225]{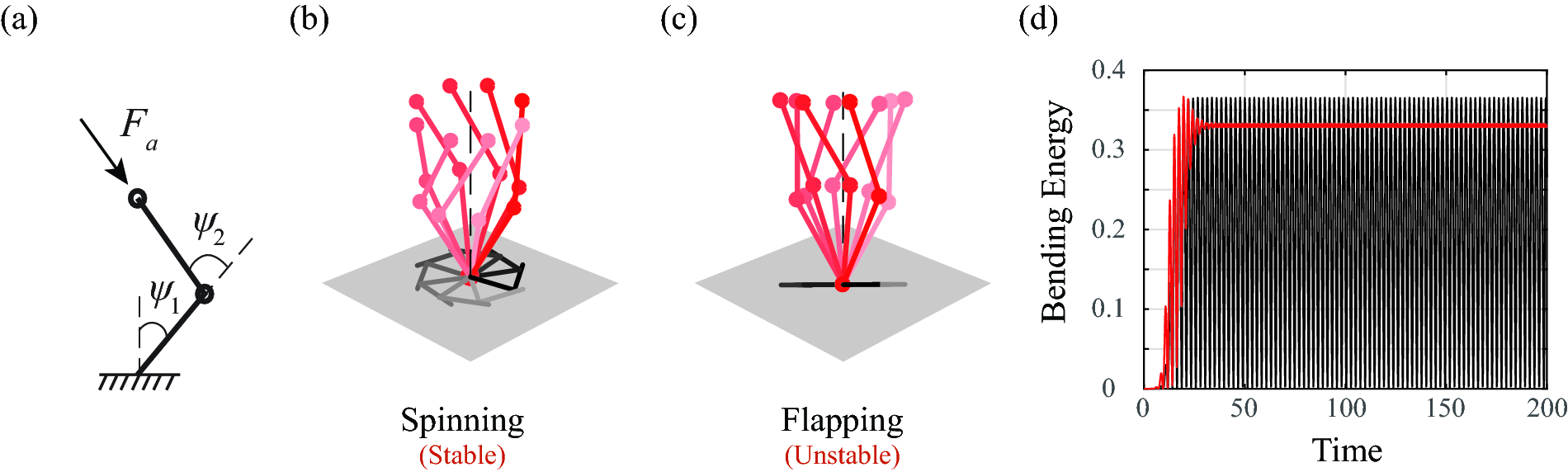}
	\caption{\textbf{Two-link chain.}  (a) A two-link chain subject to an axial follower force $F_a=4$, concentrated at its tip. (b) Starting from a straight configuration, \textit{non-planar} perturbations lead to 3D spinning motions. (c) By symmetry, \textit{Planar} perturbations lead to flapping motions, but these solutions are \textit{unstable} to the three-dimensional perturbations. (d) the maximum bending energy of planar flapping is larger than the constant bending energy of the spinning motion.}
	\label{fig:2links}
\end{figure*}

\section{Bead-spring model}
\label{sec:beads}

The elastic filament, subject to an axial force at the free tip, buckles and undergoes 3D rigid spinning at a locked curvature before it transitions to 2D flapping motions at large values of $F_a$.  
These nonlinear effects are not captured by the  linear stability analysis. Here, we consider a bead-spring  model of the filament with the aim of reproducing these nonlinear effects with the minimum number of degrees of freedom. 

We model the elastic filament as a chain of $N$ beads located at $\mathbf{r}_{i}$, where $i=1,\ldots,N$, and connected successively via inextensible rods of length $L/N$ from the origin $\mathbf{r}_o = (0,0,0)^\mathsf{\! T}$; see figure~\ref{fig:2links} for a two-link example. To emulate the filament elasticity we attach to each bead  $i\neq N$ a torsional spring, that produces a spring moment $M_{s,i}=-K\psi_i$ where $K$ is the spring constant and $\psi_i:=\cos^{-1}(\mathbf{t}_{i-1}\cdot\mathbf{t}_{i})$ is the the angle between successive links, with $\mathbf{t}_{i}:=(\mathbf{r}_{i}-\mathbf{r}_{i-1})/\|\mathbf{r}_{i}-\mathbf{r}_{i-1}\|$.
We enforce the clamped boundary condition at $\mathbf{r}_o$ weakly by including a spring moment $M_{s,o}=-K\cos^{-1}(\bm{e}_3\cdot \mathbf{t}_{1})$, proportional to the angle between the equilibrium axis and the first link. The chain is subject to an axial force acting at its tip, on bead $N$, of the form $\mathbf{F}_a=-F_a\mathbf{t}_N$. Each bead is subject to an \textit{isotropic} drag force $\mathbf{F}_{h,i}=-\zeta\mathbf{v}_{i}$.

\begin{figure}[!htb]
	\centering
	\includegraphics[scale=0.225]{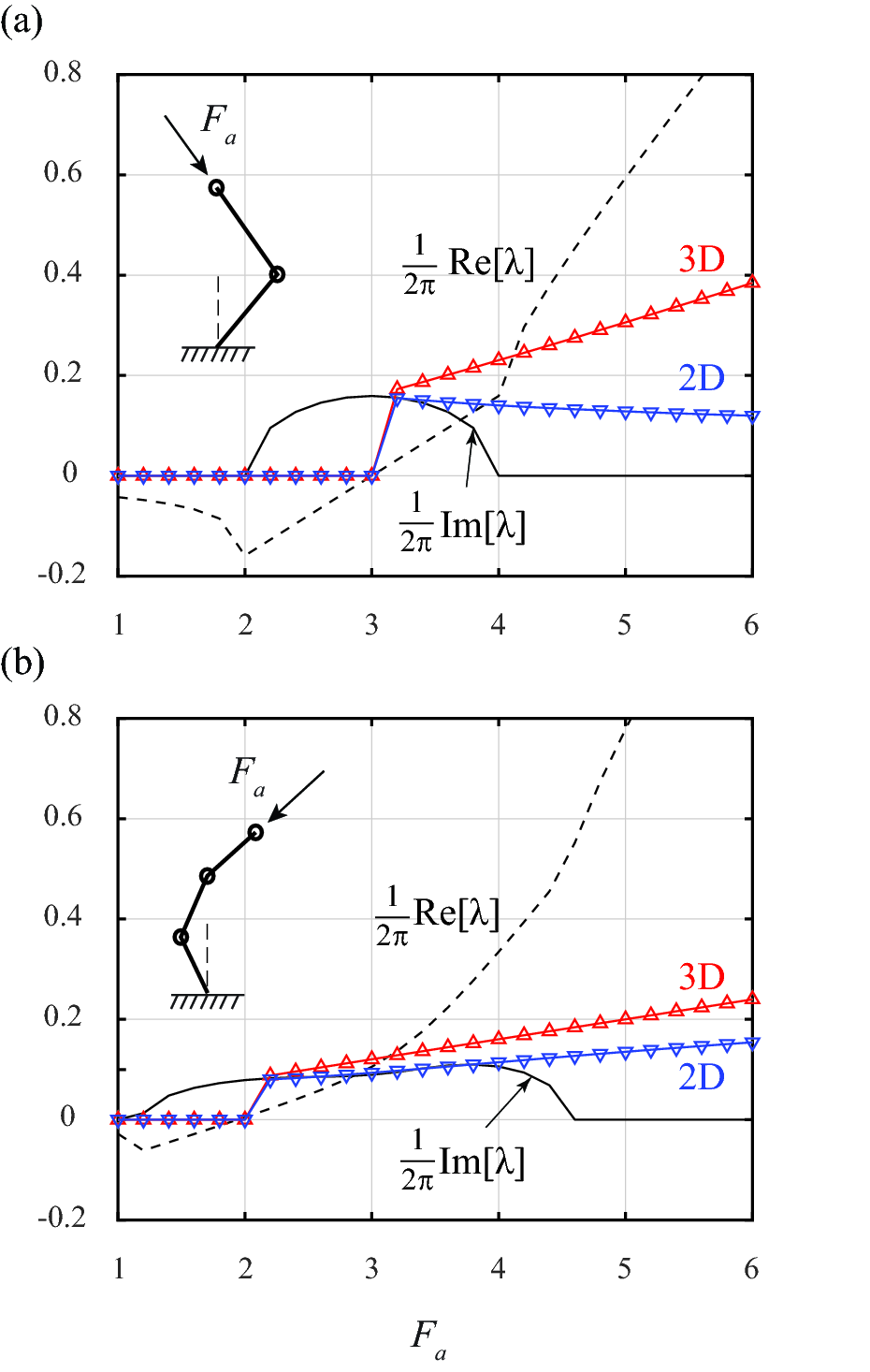}
	\caption{\textbf{Linear stability of two-link and three-link chains.} Nonlinear frequencies of the (a) two- and (b) three-link chains compared to the dominant eigenvalue from linear stability analysis. Similar to figure~\ref{fig:transitionF}, the frequency of 3D spinning is consistently larger than the frequency of planar flapping. The latter is unstable to non-planar perturbations. }
	\label{fig:linkLSA}
\end{figure}

\begin{figure*}[!t]
	\centering
	\includegraphics[scale=0.225]{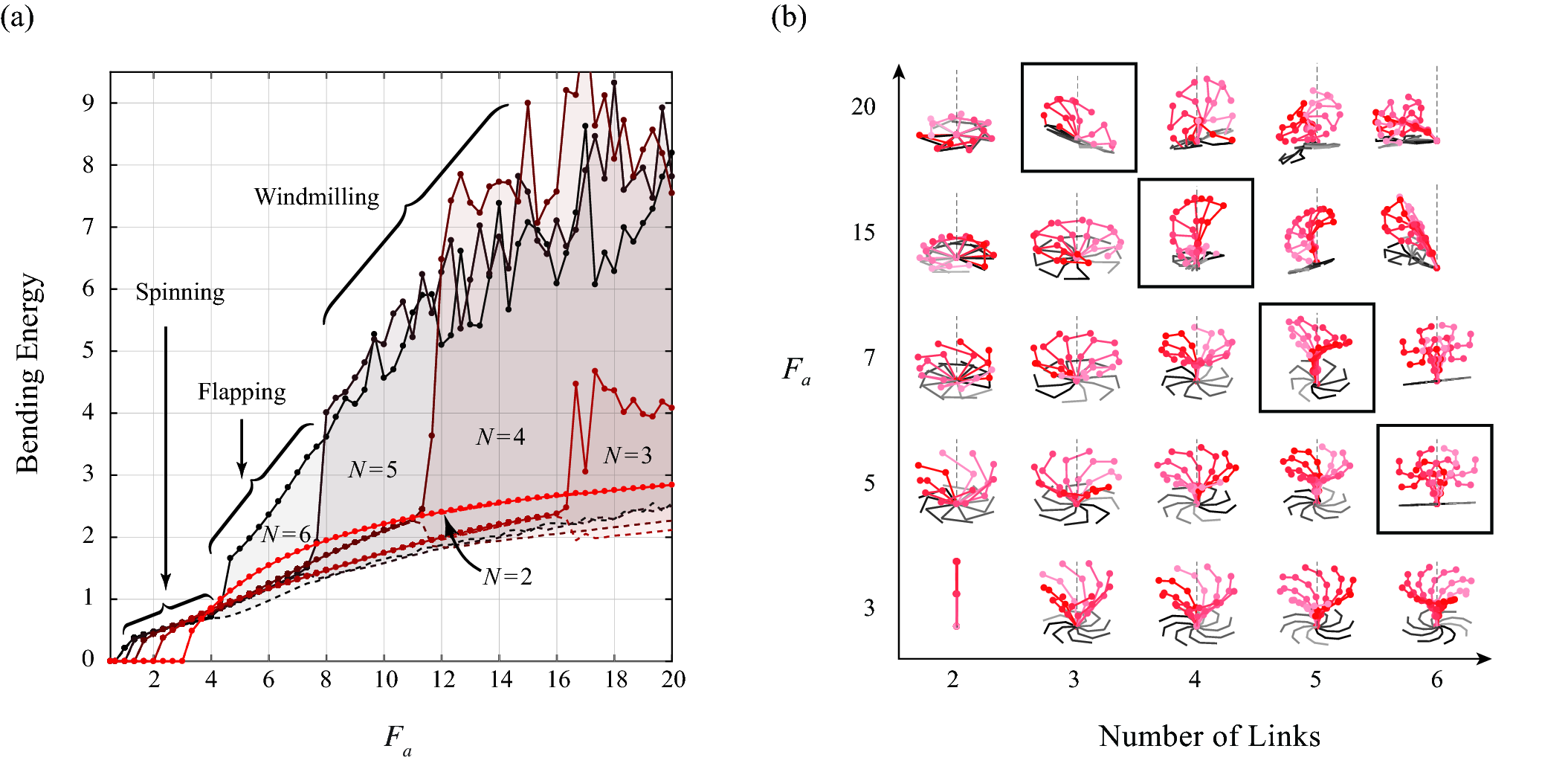}
	\caption{\textbf{Transition to non-spinning solutions of the bead-spring models.} We perform a continuation type analysis for different number of links, where we let the chain settle onto a stable solution at a given $F_a$ and successively increase $F_a$ to obtain new stable solutions. (a) The maximum and minimum envelope of the resultant bending energy as a function of $F_a$. Spinning motions become unstable at large $F_a$ for $N\geq3$. (b) Chain behavior as a function of the number of links and  $F_a$. flapping solutions become stable for a fixed range of $F_a$  only at $N=6$. }
	\label{fig:linkTrans}
\end{figure*}

We rewrite these quantities in non-dimensional form using the total chain length $L$ as the characteristic length scale and $L^3\zeta/K$ as the characteristic time scale. To this end, the non-dimensional force is given by $F_a L / K$ as opposed to the non-dimensional force in the filament model $F_a L^2/B$. Thus,  the range of relevant values of the dimensionless force in the chain model may differ from that of the filament model.  In the following, we consider all quantities are non-dimensional.

We write the position vector to bead $i$ as $\mathbf{r}_{i}=\mathbf{r}_{i-1}+\mathbf{t}_{i}/N$, where the vector $\mathbf{t}_{i}$ is expressed in terms of the spherical coordinates $\theta_i$ and $\phi_i$ attached to bead $i$, namely, $\mathbf{t}_{i}\equiv (\cos \theta_i \cos \phi_i, \cos \theta_i \sin \phi_i, \sin \theta_i)^\mathsf{\! T}$. 
We can then explicitly express the variations $\delta \mathbf{r}_i$ and use the principle of virtual work, 
\begin{equation}
\mathbf{F}_{a}\cdot\delta \mathbf{r}_{N} + \sum_{i=1}^N \mathbf{F}_{h,i}\cdot\delta \mathbf{r}_{i} + \sum_{i=0}^{N-1} M_{s,i}\delta \psi_i =0,
\end{equation}
to obtain $2N$ governing equations
for an $N$-link chain in terms of $\theta_i$ and $\phi_i$. 
Note that after proper nondimensionalization, the planar two-link equations of \cite{DeCanio2017} can be recovered by setting $\phi_i=0$, $i=0,1$. 
For $N=1$, inextensibility dictates that the tip follower force has no effect, thus the straight equilibrium configuration is globally asymptotically stable.
The system admits both spinning and flapping trajectories for all $N\geq 2$.  However, since torsion is related to $\partial^3\mathbf{r}/\partial s^3$, to account for the effects of torsion, we need $N\geq 3$.

In figure~\ref{fig:2links}, we show the behavior of a two-link model subject to an axial follower force 
$F_a=4$, concentrated at its tip. Starting from a straight configuration, \textit{non-planar} perturbations bring the chain into 3D spinning motion. Planar flapping motions can only be obtained for  \textit{planar} initial perturbations, and are \textit{unstable} to non-planar perturbations, consistent with the full filament model. The bending energy of spinning is constant over time while the bending energy of the flapping motion oscillates as a function of time, accessing larger energy values than the spinning motion, similar to the full filament model. 

We linearize the equations of motion for a two-link chain and a three-link chain and examine the dominant eigenvalues of the linear system.  The state variables $\theta_i$ from $\phi_i$ decouple at the linear level, similar to the decoupling of curvature and torsion in the full filament model.  
Figure~\ref{fig:linkLSA} shows the nonlinear frequencies of the two- and three-link chains, subject to 3D and 2D perturbations, and compares them to the results of the linear stability analysis. As in the full filament model, a Hopf bifurcation leads to sustained oscillations of the chains, and as in  figure~\ref{fig:transitionF}, the frequencies of 3D spinning are consistently larger than the frequencies of planar flapping of the confined chain. The latter solutions are unstable to non-planar perturbations.

In order to examine the transition from 3D spinning to 2D flapping motions in the chain model, we
perform a continuation type analysis for five different link numbers $N=2$ to $6$. We stop at $N=6$ because it corresponds to the chain with the smallest number of links that unambiguously  exhibit the transition from 3D spinning to 2D flapping. For each $N$, we start with $F_a=0$ and increase its value systematically as follows: we let the chain settle into a stable solution at a given value of $F_a$, we then increase the value of $F_a$ to obtain a new stable solution, and repeat successively. In figure~\ref{fig:linkTrans}(a) we show the maximum and minimum envelope of the bending energy as a function of $F_a$. Since the 3D spinning motions correspond to rigid rotations at a locked curvature of the buckled filament, the bending energy is constant, and its maximum and minimum are equal. Observe that spinning become unstable at large values of $F_a$ only for $N\geq3$. 
In figure~\ref{fig:linkTrans}(b), we illustrate the chain behavior by superimposing several snapshots of the same chain during one period of its oscillations.  We show the filament behavior as a function of the number of links $N$ and the tip force $F_a$.  
For $3 \leq N<6$ we observe a transition from spinning to `windmilling' motions, where the chain or part of the chain rotates (windmills) roughly about an axis in the $(x,y)$ plane. For $N=6$, planar flapping solutions become stable for a range of $F_a$.
{This is also true for $N>6$ (results not shown for brevity). In the limit of large $N$, the discrete chain model exhibits similar results to the discrete filament model of Section \ref{sec:num}, albeit without drag anisotropy. 
That is to say, instability-driven oscillations occur under the assumption of local isotropic drag, indicating that they are insensitive to the hydrodynamic force model. 
This is in contrast to systems that require hydrodynamic interactions along the chain to trigger sustained oscillations~\cite{Laskar2013}.
}

The transition away from the 3D spinning motion is consistent with our physical intuition in \S\ref{sec:tipforce}. When spinning, the work done by the tip force is balanced by the constant elastic energy stored in the locked configuration of the chain and the work dissipated through viscous drag due to the ``rigid'' rotation about the $z$-axis. As the tip force $F_a$ increases, to maintain this balance, the three-link chain opts to minimize time-dependent deformations by rigidly rotating the whole chain about an axis in the $(x,y)$ plane. This rotation is permitted because the clamped boundary condition in the chain model is imposed weakly, via an elastic spring at $O$. Since these rotations activate the elastic spring at the base, the corresponding elastic energy is non-zero. 
Also, these rotations lead to larger linear velocities and thus larger drag forces (that is, larger dissipation) than rotations about the $z$-axis. 
For four- and five-link chains, the upper link of the chain succumbs first under the axial force and begins to rotate in a windmill-like motion, as illustrated in figure~\ref{fig:linkTrans}(b).
However, as the number of links increase, these hybrid solutions -- consisting of rigid spinning about the $z$-axis with windmill-like rotations that are spatially-localized at the distal links -- become unstable. At $N=6$, the chain clearly exhibits wave-like deformations.

\section{Discussion}
\label{sec:discussion}

We considered an elastic microfilament, clamped at one end in a viscous fluid, subject to a  distribution of axial forces. We identified two major transitions in the filament behavior.
First, the filament transitions from a straight configuration to a buckled configuration, with locked curvature, and undergoes 3D spinning. The second transition occurs at larger values of the axial force and marks a destabilization of the 3D spinning motions, giving rise to 2D flapping oscillations.  
These periodic motions, both 3D spinning and 2D flapping, are robust to large perturbations in the filament configuration.
They are also robust, away from the bifurcation values, to perturbations in the parameters of the axial forces, and even to changes in the force profile altogether  (as long as the force density increases sufficiently nonlinearly towards the tip of the filament).
These findings, {although in the context of a simplified model,} support the idea that an open-loop, instability-driven mechanism could explain not only the origin of sustained oscillations but also the wide variety of periodic beating patterns observed in cilia and flagella.

\begin{figure*}[!htb]
	\centering
	\includegraphics[scale=2.4]{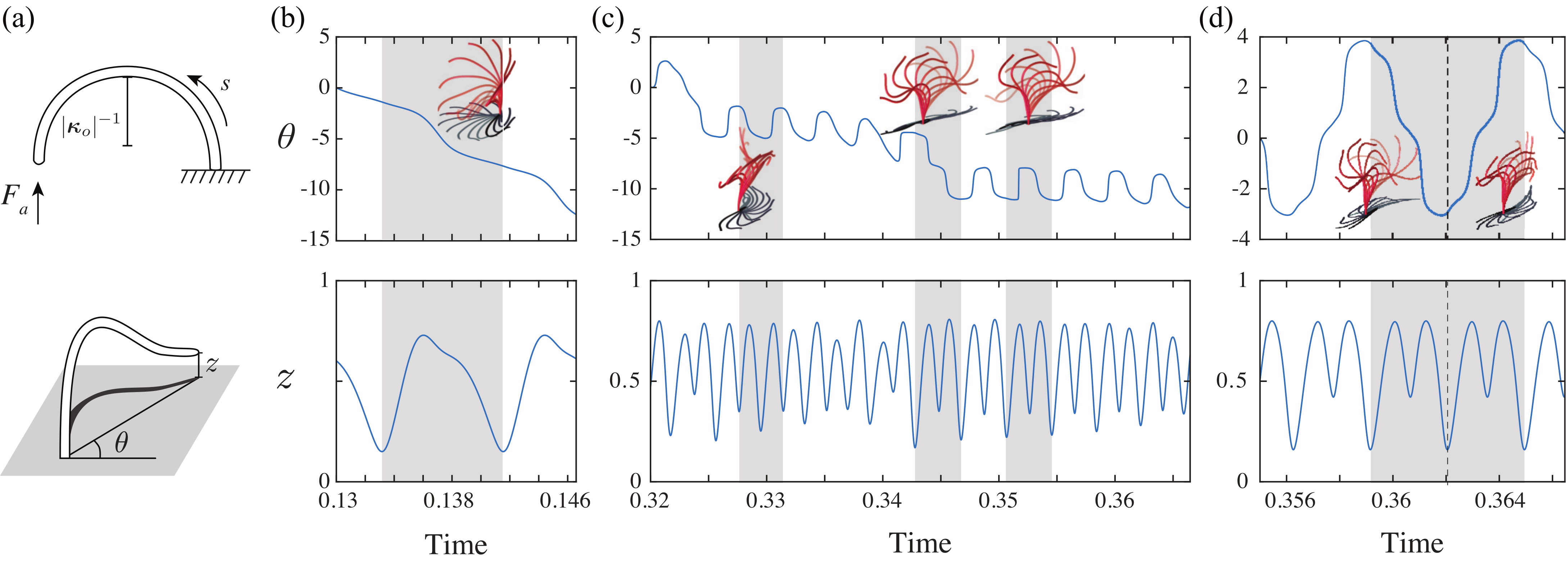}
	\caption{\textbf{Asymmetric beating patterns.} (a) A naturally-curved filament with a reference curvature of $\boldsymbol{\kappa}_o= 2\pi\mathbf{d}_1$ is subjected to an axial force $F_a$ at its tip. We define the rotation angle $\theta$ in the $(x,y)$-plane associated with the vertical projection of the tip point. (b) At $F_a = 100$, an asymmetric 3D spinning pattern develops: $\theta$ monotonically decreases and the vertical tip position $z$ varies periodically. (c) At $F_a = 200$, the filament engages in asymmetric flapping while it slowly precesses about the vertical axis: $\theta$ drifts on average. (d) At $F_a = 250$, the asymmetric flapping is nearly planar but switches between flapping on the left and right sides of that plane: $\theta$ is periodic. The filament snapshots are taken from the regions highlighted in grey.}
	\label{fig:staticK}
\end{figure*}

Cilia and flagella, despite their highly conserved internal axoneme structure across cell types, exhibit distinct beating patterns depending on the cell type, environment and function.
For example, sperm cells propel themselves by planar wave-like deformations of their flagellum~\cite{Woolley2009}. 
The biflagellate algal cell \textit{Chlamydomonas} swim using synchronized flapping oscillations of their flagella, referred to as  `breast stroke,' and can transition to unsynchronized wave-like oscillations, referred to as `free style,' that allows them to turn and reorient themselves~\cite{Leptos2013, Wan2014}.
In densely-packed ciliated surface, motile cilia often beat in 3D patterns, with an asymmetry in their effective and recovery stroke. A gallery of various planar and non-planar cilia beating patterns can be found in~\cite{Sleigh1968,Brennen1977}.  

{The model presented in this paper does not accurately represent the intricate internal structure of cilia. It rather assumes that its net effect on the cilium centerline can be approximated by a distribution of axial forces of constant magnitude. An accurate representation of the effect of the axoneme structure on the centerline motion would give rise to a distribution of axial and normal forces that are coupled to the centerline geometry. These limitations notwithstanding, we use the current model to roughly gauge whether the instability-driven oscillations and the transitions reported in this study are relevant to the mechanics of cilia and flagella.} To this end, we provide a rough estimate of the total axial force that is biologically attainable. 
We estimate the total amount of forces that the dynein motors provide inside a typical cilium. The axoneme consists typically of `9+2' microtubule doublets, with nine outer doublets and a central pair (see figure~\ref{fig:cilia}). Dynein molecules form an array of cross-bridges between pairs of outer microtubule doublets.
The force exerted on microtubules by a single molecular motors is known to be of the order of a few pico-newtons. So assuming 8 dynein arms per 100 nm and 3-8 pN of force per dynein arm \cite{Shingyoji1998}, we would get a per-doublet force density estimate of 200-600 pN$\cdot\mu$m$^{-1}$.This gives us about 2-5 nN$\cdot\mu$m$^{-1}$ per cilium assuming all 9 doublets are activated. 
In reality, due to the internal structural damping and other reaction forces (most doublets experience dynein forces in \textit{both} compressive and extensile directions), the total compressive force density is likely to be much smaller.
In our model, a force density on the order of 20 pN$\cdot\mu$m$^{-1}$ can lead to sustained oscillations and is sufficient to even induce the transition from 3D spinning to planar flapping regime.
This calculation is based on the cilia length and bending rigidity estimate of Table \ref{tab:dim}, where we have chosen a typical length of 20 $\mu$m as cilia and flagella can vary in length between 10-50 $\mu$m, and a high stiffness value of 800 pN$\cdot\mu$m$^2$ comparing to values of 25-800 pN$\cdot\mu$m$^2$ reported in literature~\cite{Eloy2012, Xu2016, Guo2014, Guo2016a, Guo2018}. 
These numbers suggest that the axial forces due to dynein activity in cilia and flagella should be large enough to support these instability-driven oscillations, as noted in more elaborate models of \cite{Bayly2016,Hu2018,Han2018}.

Another important consideration for exploring the relevance of these instability-driven oscillations to biological cilia is that cilia often exhibit asymmetric beating patterns. To account for asymmetric oscillations in our filament model, we incorporate a non-zero reference curvature of the filament.
Experimental evidence from the algal cells \textit{Chlamydomonas} supports the notion that their flagella are naturally curved~\cite{Sartori2016a}.
In our model, we set $\mathbf{M}=\mathbf{B}(\boldsymbol{\kappa}-\boldsymbol{\kappa}_o)$, where $\boldsymbol{\kappa}_o$ is a non-zero reference curvature. We consider $\boldsymbol{\kappa}_o=2\pi\mathbf{d}_1$, as depicted schematically in the top of Figure \ref{fig:staticK}(a).
Figure \ref{fig:staticK} shows the filament dynamics under three values of the axial force $F_a$ at the free tip of the filament.
We identify three distinct behaviors of the naturally curved filament: (i)  ``asymmetric'' spinning (figure \ref{fig:staticK}b), reminiscent to the spinning behavior of the naturally straight filament, but here the spinning axis is not normal to the base wall to which the filament is clamped; (ii) ``asymmetric'' precession while flapping (figure \ref{fig:staticK}c), where the flapping motion is combined with a slow precession about the vertical axis, and flapping frequency about five times faster than the precession frequency; and (iii) ``asymmetric'' flapping  (figure \ref{fig:staticK}d) that is nearly, but not completely planar; the filament switches between the left and right sides of a flapping plane in an alternating fashion, as evident from the shadows in the inset of Figure~\ref{fig:staticK}d.
These regimes can be cleanly delineated from the time evolution of a cumulative rotation angle $\theta$, defined in the $(x,y)$-plane as the angle to the vertical projection of the filament tip, see the bottom schematic in figure \ref{fig:staticK}(a).
In the top row of figures~\ref{fig:staticK}(b-d), we plot the change in $\theta$ after the filament reaches its limit cycle behavior. At $F_a=100$ (figure~\ref{fig:staticK}b), $\theta$ change monotonic in time, indicating that the filament is simply spinning. At $F_a=200$ (figure~\ref{fig:staticK}c), the fast periodic fluctuation of $\theta$ reflects the flapping motion, and the slow drift of its average value captures the filament precession. At $F_a=250$ (figure~\ref{fig:staticK}c), $\theta$ varies in $(-\pi,\pi]$ indicating that the motion is purely flapping. Due to the asymmetry incorporated in the system, the vertical position $z$ for the filament tip changes with time in all three behaviors; see bottom row of figures~\ref{fig:staticK}(b-d). In particular, over one period of oscillation (some highlighted in grey in figure~\ref{fig:staticK}), the filament tip periodically gets closer to then further away from the base wall, in a way reminiscent to the effective and recovery strokes in ciliary beating patterns.

The work in this study is a first step towards understanding the interplay between the molecular motor activity, the elastic properties of biological filaments, and the fluid mechanical forces in generating the wide variety of beating patterns observed across flagellated cells and in the transitions between multiple beating modes in the same cell as the level of molecular motor activity changes.  This work focused on the case of a single filament in a viscous fluid, using the resistive force theory. 
One of the main advantages of this simple model is that it avoids the requirement of numerous  material parameters, unknown experimentally, such as those needed in a full-fledged finite element simulation involving more details of the internal structure of flagella \cite{Hu2018,Chen2015}.
This simple model also allows us to focus on the fundamental mechanisms underlying flagellar oscillations and the various transitions that could potentially be responsible for the variety of beating patterns observed in nature.
However, a few remarks on the limitations of the model are in order.
Our model does not fully reflect the internal mechanics of cilia. In ongoing work,
we are deriving the equations of motion and effective forces acting on the centerline from more elaborate, coupled, multi-filament models of the axoneme structure.
The present model does not account for the long range hydrodynamic effects of viscous fluids. 
Future extensions of this work will account for the full hydrodynamics, using a combination of slender body theory and regularized Stokeslet formulation, in the presence of the cell body or the cell wall~\cite{Wrobel2016, Guo2018}.
Future work will also focus on the effects of anisotropy in the bending rigidity matrix $\mathbf{B}$, which encodes information about the internal axoneme structure. Experimental evidence indicates that the presence or absence of certain structures within the flagellum could lead to 3D versus planar beating patterns in the \textit{Chlamydomonas} flagella~\cite{Meng2014}. This suggests that anisotropy of bending rigidity could be another crucial effect responsible for the varied motion of cilia and flagella. It is also important to account for the effects of the intrinsic oscillations of the dynein motor forces due to the time scales of their binding and unbinding to microtubules and due to thermal fluctuations~\cite{Shingyoji1998}. Lastly, this instability-driven mechanism could have implications on the coordinations of multiple active filaments with non-local hydrodynamic interactions~\cite{Guo2018}. The presence of neighboring filaments and other solid boundaries may affect the transition thresholds and bias planar versus non-planar beating patterns.

\bibliographystyle{vancouver}
\bibliography{references}

\end{document}